\documentclass[sigconf]{acmart}
\AtBeginDocument{%
  }

\setcopyright{acmlicensed}
\copyrightyear{2026}
\acmYear{2026}
\setcopyright{cc}
\setcctype{by}
\acmConference[KDD '26]{Proceedings of the 32nd ACM SIGKDD Conference on Knowledge Discovery and Data Mining V.2}{August 09--13, 2026}{Jeju Island, Republic of Korea}
\acmBooktitle{Proceedings of the 32nd ACM SIGKDD Conference on Knowledge Discovery and Data Mining V.2 (KDD '26), August 09--13, 2026, Jeju Island, Republic of Korea}
\acmDOI{10.1145/3770855.3818458}
\acmISBN{979-8-4007-2259-2/2026/08}

\newcommand{\algns}{$\mathsf{Optimus}$}
\newcommand{\alg}{$\mathsf{Optimus}$~}
\usepackage{listings}
\usepackage{multirow}

\usepackage[utf8]{inputenc}
\usepackage{geometry}
\usepackage{array}
\usepackage{graphicx}
\usepackage{subcaption} % Required for subfigure environment

\usepackage[ruled,vlined,linesnumbered]{algorithm2e}

\newcolumntype{M}[1]{>{\centering\arraybackslash}m{#1}}
\begin{document}
\title{\algns: A Generic Operator-Level PyTorch Model Transformation Framework}

\author{Menglu Yu}
\affiliation{%
  \institution{Meta}
  \city{Bellevue}
  % \state{WA}
  \country{USA}}
\email{mengluy@meta.com}

\author{Jiaqi Xu}
\affiliation{%
% \department{Ads ML Infra}
  \institution{Meta}
  \city{Sunnyvale}
  % \state{CA}
  \country{USA}}
\email{jackiexu0313@meta.com}

\author{Yuzhen Huang}
\affiliation{%
  \institution{Meta}
  \city{Sunnyvale}
  \country{USA}}
\email{yuzhenhuang@meta.com}

\author{Yanbo Liang}
\affiliation{%
  \institution{Meta}
  \city{Menlo Park}
  \country{USA}}
\email{ybliang@meta.com}

\author{Jia Liu}
\affiliation{%
  \institution{The Ohio State University}
  \city{Columbus}
  \country{USA}}
\email{liu@ece.osu.edu}

\author{Shuai Yang}
\affiliation{%
  \institution{Meta}
  \city{Sunnyvale}
  \country{USA}}
\email{shuaiyang@meta.com}

\author{Jason Ansel}
\affiliation{%
  \institution{Meta}
  \city{Menlo Park}
  \country{USA}}
\email{jansel@meta.com}

\author{Elias Ellison}
\affiliation{%
  \institution{Meta}
  \city{New York}
  \country{USA}}
\email{eellison@meta.com}

\author{Edward Yang}
\affiliation{%
  \institution{Meta}
  \city{New York}
  \country{USA}}
\email{ezyang@meta.com}

\author{Brian Hirsh}
\affiliation{%
  \institution{Meta}
  \city{New York}
  \country{USA}}
\email{hirsheybar@meta.com}

\author{Jia Chen Ren}
\affiliation{%
  \institution{Meta}
  \city{Menlo Park}
  \country{USA}}
\email{bobren@meta.com}

\author{Will Feng}
\affiliation{%
  \institution{Meta}
  \city{Menlo Park}
  \country{USA}}
\email{willfeng@meta.com}

\author{Oguz Ulgen}
\affiliation{%
  \institution{Meta}
  \city{Menlo Park}
  \country{USA}}
\email{oulgen@meta.com}

\author{Xu Zhao}
\affiliation{%
  \institution{Meta}
  \city{Toronto}
  \country{Canada}}
\email{xzhao9@meta.com}

\author{Daohang Shi}
\affiliation{%
  \institution{Meta}
  \city{Bellevue}
  \country{USA}}
\email{daohang@meta.com}

\author{Huaqing Xiong}
\affiliation{%
  \institution{Meta}
  \city{New York}
  \country{USA}}
\email{huaqingxiong@meta.com}

\author{Quanyu Zhu}
\affiliation{%
  \institution{Meta}
  \city{Sunnyvale}
  \country{USA}}
\email{qyz@meta.com}

\author{Mingming Ding}
\affiliation{%
  \institution{Meta}
  \city{Bellevue}
  \country{USA}}
\email{midin@meta.com}

\author{Junqing Zhou}
\affiliation{%
  \institution{Meta}
  \city{Sunnyvale}
  \country{USA}}
\email{junqingz@meta.com}

\author{Ruilin Chen}
\affiliation{%
  \institution{Meta}
  \city{Bellevue}
  \country{USA}}
\email{ruilinchen@meta.com}

\author{Yuhang Yang}
\affiliation{%
  \institution{Meta}
  \city{Bellevue}
  \country{USA}}
\email{yuhangyang@meta.com}

\author{Chi-Keung Luk}
\affiliation{%
  \institution{Meta}
  \city{Sunnyvale}
  \country{USA}}
\email{kluk@meta.com}

\renewcommand{\shortauthors}{Menglu Yu et al.}

\begin{abstract}
In large-scale industrial applications, deep learning models that power recommendation and ranking have complex and diverse model architectures. 
These models are continuously developed and refined by large teams of machine learning engineers, rendering manual optimization infeasible. 
% Consequently, graph-based optimization techniques that most notably those implemented via PyTorch FX transformations have become a standard practice for performance optimization.
Consequently, graph-based optimization techniques have become an industry standard for boosting performance, with PyTorch FX transformations leading the charge.
These transformations typically rely on a set of human-engineered module-level rewrite rules which are not scalable to diverse model architectures. 
%Graph transformations, such as those enabled by FX transformation, play a critical role in enhancing model performance within deep learning frameworks.
%Although the effectiveness of such transformations for the large language models (LLMs) is still under exploration, their impact on recommendation models with diverse model architectures has proven substantial.
% These transformations typically rely on a set of module-level rewrite rules that are manually designed. However, such human-engineered rules are often tightly coupled to specific modules, limiting their scalability and generalizability across diverse model architectures.
To address this limitation, we introduce \algns, a general-purpose model transformation framework built in the PyTorch 2.x (PT2) machine learning compiler. 
% Leveraging PT2’s ability to convert models into graph representations, \alg enables efficient and customizable operator-level graph transformations without requiring changes to the original PyTorch code.
% PT2 provides multiple intermediate representations—such as torch, aten, and kernel—that serve as suitable entry points for a wide range of optimization strategies. \alg builds upon these by offering a collection of atomic graph transformation rules. Users can configure both the rules and their execution order, allowing for highly customizable transformation pipelines tailored to specific model optimization needs.
With a concise set of predefined patterns, \alg applies an efficient greedy search algorithm for pattern matching and replacement, while preserving model semantic. It is designed and implemented as a highly customizable and extensible framework integrated into the PT2 stack.
% Specifically, for a given compiled model graph, \alg  applies an efficient greedy search algorithm to match the patterns specified in each rule. The framework then rewrites the graph accordingly, ensuring that the original model semantics are preserved throughout the transformation process.
Our evaluation shows that the framework can achieve up to 63\% speedup, 6\% peak memory reduction, and over $400$ second compile time decrease for our industry-scale recommendation models compared to baselines. 
\alg is open-sourced together with PyTorch 2.x as a customizable model transformation layer.
\end{abstract}





\maketitle

% Define distinct macros for each artifact URL
\newcommand\kddcodeurl{https://doi.org/10.5281/zenodo.20422163}
\newcommand\kdddataurl{https://doi.org/10.5281/zenodo.20422171}

\begingroup\small\noindent\raggedright\textbf{Resource Availability:}\\
The artifacts of this paper have been made publicly available. The source code is accessible at \url{\kddcodeurl}, and the illustrative example model can be found at \url{\kdddataurl}.
\endgroup

\section{Introduction}
Over the past decade, Deep Learning (DL) has achieved remarkable progress, largely driven by the development of increasingly large models and complex architectures that require significant computational and memory resources~\cite{dean2012largescale}.
A central challenge in this domain lies in balancing i) the need for model complexity that is often essential to achieve higher accuracy, with ii) the practical constraints of deployment environments, where power efficiency and limited hardware resources are critical considerations.
To address this challenge, recent research has focused on optimizing DL models through {\em graph transformations,} where models are represented as tensor computation graphs~\cite{chen2018tvm}.
More specifically, given a tensor computation graph $G$ and a set of semantics-preserving transformation rules $R$, the goal of graph transformation is to identify subgraphs in $G$ that match patterns in $R$ and rewrite them with semantic-equivalent subgraphs to optimize for a specific performance objective.
% In real-world ranking and recommendation application, model architectures are diverse and are continuously evolving.
% Manual optimizations are model-specific and do not scale to the large number of model variants used in production and experimentation.
% Recent research and practice has focused on optimizing DL models through {\em graph transformations}, where models are represented as tensor computation graphs~\cite{chen2018tvm}.
% More specifically, given a tensor computation graph $G$ and a set of semantics-preserving transformation rules $R$, the goal of graph transformation is to identify subgraphs in $G$ that match patterns in $R$ and rewrite them with semantic-equivalent subgraphs to optimize for a specific performance objective.

% \begin{figure*}[t!]
%     \centering
%     \begin{minipage}[t]{0.45\textwidth}
%         \centering
%         \includegraphics[width=\linewidth]{pictures/PT2 overview.png}
%         \caption{PyTorch 2.x stack overview.}
%         \label{fig:PT2_overview}
%     \end{minipage}
%     \hfill
%     \begin{minipage}[t]{0.45\textwidth}
%         \centering
%         \includegraphics[width=\linewidth]{pictures/integration.png}
%         \caption{\alg backend integration points.}
%         \label{fig:integration}
%     \end{minipage}
% \end{figure*}

Modern ML compilers, such as PyTorch 2.x (PT2)~\cite{Ansel2024pt2}, preserve the familiar eager-mode user experience of PyTorch while introducing powerful compiler-level optimizations.
Using symbolic tracing, PT2 converts models into graph representations, enabling more effective manipulation and transformation.
PT2 supports multiple levels of graph intermediate representation (IR), such as Torch, ATen, and Inductor IR, each facilitating various graph optimization strategies.
The first attempt is to employ FX-based graph transformations~\cite{nie2022optimizing}, which mainly focuses on module level.
Although effective in some scenarios, this module-level approach can be inefficient,
%than operator-level fusion, 
since it often fails to capture and fuse numerous fine-grained operations, thus leading to suboptimal GPU utilization.
Moreover, the rapid evolution of DL model architectures and frequent modifications to modules present scalability challenges for module-level fusion, limiting its applicability in dynamic or production-scale settings.
To address the scalability limitations at the module level, subsequent research focused on enabling more fine-grained, operator-level model transformations~\cite{xu2023empowering}.
A key challenge in this space is defining generalizable transformation rules to replace heuristic-driven model optimizations.
Several research efforts have explored automated optimization techniques.
For example, TASO~\cite{jia2019taso} automates graph substitutions for new operators.
OCGGS~\cite{fang2020optimizing} introduces a method for efficiently determining optimal execution times; and TENSAT~\cite{yang2021equality} applies equality saturation to decouple optimization into two phases.
However, these methods often suffer from high computational overhead, where subgraph substitution searches can take hours even for relatively simple models, making them impractical for real-time training scenarios or deployment in complex production pipelines.

To address the limitations of existing approaches, we observe that many human-crafted graph optimizations can be systematically decomposed into compositions of atomic operator-level fusions.
By breaking down complex transformation logic into fine-grained, operator-level patterns, our framework enables scalable and efficient model optimization without compromising flexibility or performance.
Consequently, it inherently avoids costly subgraph substitution searches while maintaining both generality and effectiveness.
Our main contributions are summarized as follows:
% \vspace{-0.15in}
\begin{itemize}
    \item We propose \algns\footnote{The source code for this work is included in PyTorch, available at \url{https://github.com/pytorch/pytorch/}. Optimus code exists in the \texttt{torch/\_inductor/fx\_passes} directory.}, an efficient general purpose operator-level graph transformation framework designed to efficiently rewrite subgraphs based on rules defined over atomic operators.
    The framework is designed to support transformation across all PT2 graph IRs, enabling reductions in execution latency, peak memory usage and compile time.
    Implemented as a PT2 plugin, \alg offers an extensible interface that allows integration of new transformation rules to meet customized optimization objectives.
    \item \alg employs an efficient greedy search algorithm to identify target subgraphs for each predefined transformation rule.
    Specifically, the search begins at designated anchor nodes and explores their neighborhoods to find desired graph patterns.
    This strategy is motivated by the observation that most atomic operator rules involve relatively small subgraph rewrites, which reduces search overhead while maintaining the flexibility to apply optimizations across a wide range of models.
    Once a matching subgraph is found, it is replaced with an optimized variant that improves performance.
    \item We extensively evaluate \alg on top five industry-scale recommendation models and show that \alg can achieve up to 63\% speedup, 6\% peak memory reduction and over 400 seconds compile time decrease compared to the baselines.
\end{itemize}

\section{Preliminaries}\label{sec:preliminaries}

In this section, we provide the necessary background relevant to \alg to acquaint readers with the terminology and context used throughout this paper.

% \begin{figure}[t!]
%     \centering
%     \includegraphics[width=.9\linewidth]{pictures/PT2 overview.png}
%     \caption{PyTorch 2.x stack overview.}
%     \label{fig:PT2_overview}
% \end{figure}

\begin{figure}[t!]
    \centering
    \includegraphics[width=.8\linewidth]{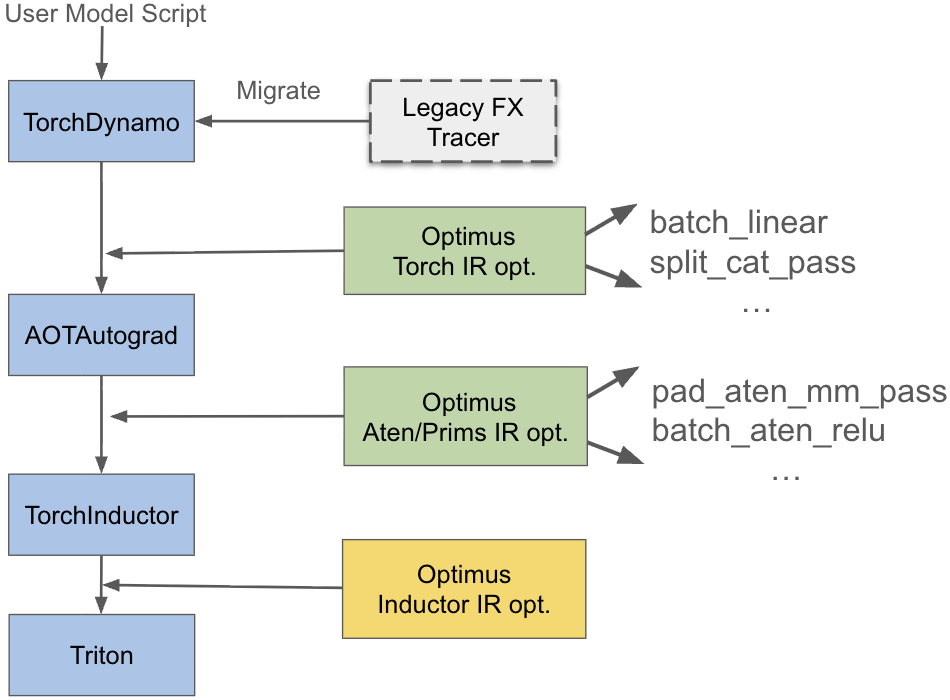}
    \caption{\alg backend integration points within PT2.}
    \label{fig:integration}
\end{figure}

\subsection{PyTorch 2.x Stack Overview}
The PT2 deep learning compiler framework offers significant performance enhancements through advanced compilation and optimization techniques.
Fig.~\ref{fig:integration} provides a high-level overview of the PT2 stack with its primary building blocks as well as how \alg integrated within the framework.
A central innovation is the \textit{torch.compile} API, which enables just-in-time model compilation, offering potential speedups for both training and inference.
This functionality is powered by several key technologies: TorchDynamo, AOTAutograd, and TorchInductor.
% TorchDynamo captures PyTorch programs as graphs, enabling static graph compilation and optimization.
% AOTAutograd replaces PyTorch’s autograd engine with a tracing-based autodiff system that produces ahead-of-time backward computation graphs.
% PrimTorch simplifies over 2,000 PyTorch operators into a closed set of approximately 250 primitive operators, reducing the complexity of developing new PyTorch features or backends. TorchInductor, the framework’s deep learning compiler, generates optimized code for a variety of accelerators and backends. For example, it uses OpenAI Triton as a key component for NVIDIA and AMD GPUs.

Specifically, under this architecture, the legacy FX Tracer~\cite{reed2022torch} has been replaced by TorchDynamo, the graph capture component responsible for generating Torch-IR graphs.
AOTAutograd dynamically records autograd logic in an ahead-of-time manner, producing distinct forward and backward computation graphs for a given model.
TorchInductor takes the ATen/Prim-based graphs produced by AOTAutograd as inputs and further lowers them to a loop-level IR.
At this stage, TorchInductor performs a set of optimizations, such as creating fusion groups to enable efficient code generation.
% TorchInductor currently supports lowering for pointwise, reduction, scatter/gather, and window operations.
% Hardware vendors can integrate with the PT2 stack by mapping the loop-level IR to hardware-specific code.
Currently, TorchInductor offers two primary backends: (1) a C++ backend that generates multithreaded CPU code, and (2) an OpenAI Triton backend for generating high-performance GPU kernels~\cite{triton2021}.
% \subsection{\alg PT 2.x Backend Integration}

After establishing an overall understanding of the PT2 stack, we now describe the \alg backend integration.
As shown in Fig.~\ref{fig:integration}, the framework supports multiple integration points, determined by the specific graph IR on which the transformation is applied.
We have implemented a variety of patterns with atomic operators across different graph IRs, with the current primary focus placed on the Inductor IR stage to optimize the generated Triton kernels.

\begin{figure}[!t]
    \centering
    \includegraphics[width=0.7\linewidth]{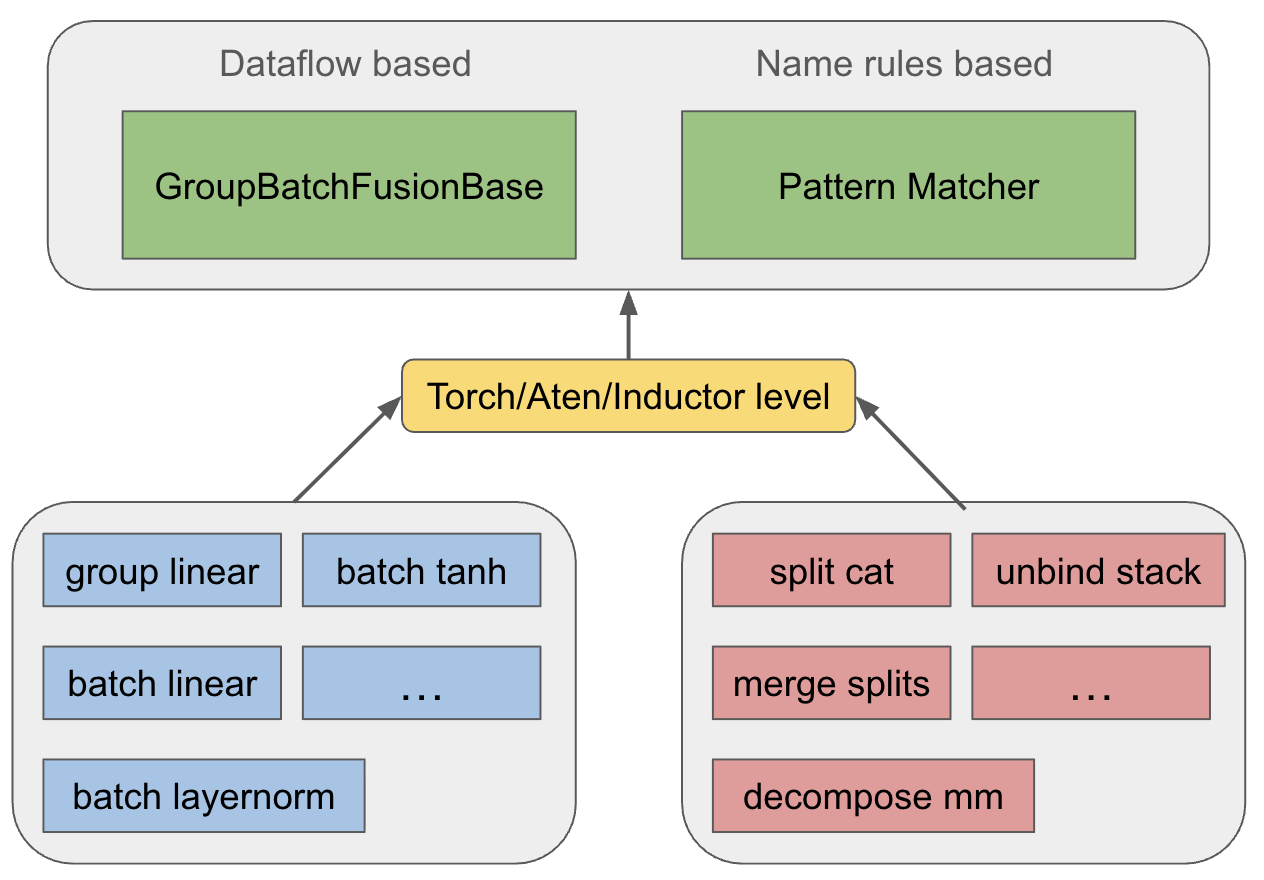}
    \caption{\alg architecture overview.}
    \Description{Overview diagram of the \alg architecture.}
    \label{fig:high-level-arch}
\end{figure}

\subsection{Semantic-Preserving Program Rewrite}
The core challenge in semantic-preserving program rewrite lies in defining effective pattern replacement rules, particularly when dealing with complex programs, while ensuring that these rules remain applicable across a wide range of program structures.
An additional challenge is the design of an efficient search algorithm, without which, the associated overhead can become substantial, potentially negating the benefits of optimization for large-scale programs, especially in latency-sensitive scenarios.

In terms of semantic-preserving program rewrite, we give the initial program $p$, where PT2 compilation produces one or more DAGs (see an example in Appendix~\ref{sec:graph-representative}), and a set of rewrite rules $R$.
Each rule has the form $r_s\rightarrow r_d$, where $r_s$ represents the original pattern in $p$, and $r_d$ is the optimized pattern.
The replacement should not change the semantic of the program, i.e., given the same input, pattern $r_s$ and $r_d$ have the same output.
The optimization starts with pattern search $r_s$ in $p$, as long as such pattern found, we will replace it with predefined pattern $r_d$.
% Given the diversity of models, graph transformations must not affect numerical correctness. 
To ensure numerical stability and correctness, We provide flexible enable and disable controls, support bisection debugging to isolate problematic patterns, and perform runtime numerical checks to verify that outputs before and after Optimus remain closely matched for identical inputs.

Note that these patterns are designed manually.
There are in general two approaches to collect these source pattern graphs: i) Analyze GPU execution traces of production models to identify ``fragmented" execution, specifically kernels with low compute intensity but high launch overhead.
We then map these inefficient kernels back to the corresponding subgraph in the PT2 graph-IR.
This allows us to identify the exact operator sequences responsible for the bottleneck; ii) Investigate the PyTorch model architecture and graph-IR to analyze operator dependencies.
By understanding the semantics of the model, we identify redundant data shuffling (such as unnecessary split/cat cycles) or opportunities for operator reordering that are semantically invariant but computationally more efficient.
Once an inefficient subgraph is identified via these two methods, we design a transformation pair: 1) the source pattern representing the current state, and 2) the optimized pattern representing the more efficient equivalent. These patterns are then added to the Optimus library, where our greedy search engine applies them recursively across other model architectures.

\section{The \alg Framework}
In this section, we 
first provide an overview of the \alg framework in Section~\ref{subsec:ArchOverview}.
We then discuss the model transformation in Section~\ref{subsec:generic-rule}, which is followed by \algns's graph and rule representations in Section~\ref{subsec:OptimusRep}.

\subsection{Architecture Overview} \label{subsec:ArchOverview}

The core components of \alg include an interface that enables users to both understand existing optimization strategies and design new model transformation rules that yield globally optimized solutions across diverse models.
Users have the flexibility to select transformation rules and determine their execution order, thereby tailoring the optimization process to maximize model performance, for example, by minimizing training latency.

As illustrated in Fig.~\ref{fig:high-level-arch}, \alg provides two complementary interfaces: a dataflow-based interface for horizontal optimizations and a rule-based interface for vertical optimizations (see Section~\ref{subsec:generic-rule} for details), which supports model transformation across multiple graph IRs within the PT2 stack.
For instance, horizontal optimizations can be applied to operators, such as linear and layer normalization, while vertical optimizations can capture structured patterns such as split–concat transformations.

\alg introduces a refined methodology for model optimization by decomposing complex heuristics into modular, atomic optimization tasks.
Each task represents a concrete instance of model transformation, which modifies the model's IR to improve training or serving efficiency without sacrificing accuracy.
Crucially, these transformations are broadly applicable across different IR levels, including torch-level, aten-level, and even inductor-level representations.
Since the transformation process follows common procedures, \alg offers a unified interface to define, apply, and manage optimizations.
This streamlined design not only simplifies the development of customized optimization strategies, but also enhances transparency and debuggability for individual tasks.

\subsection{Model Transformation}\label{subsec:generic-rule}
\begin{figure}[!t]
    \centering
    \includegraphics[width=0.8\linewidth]{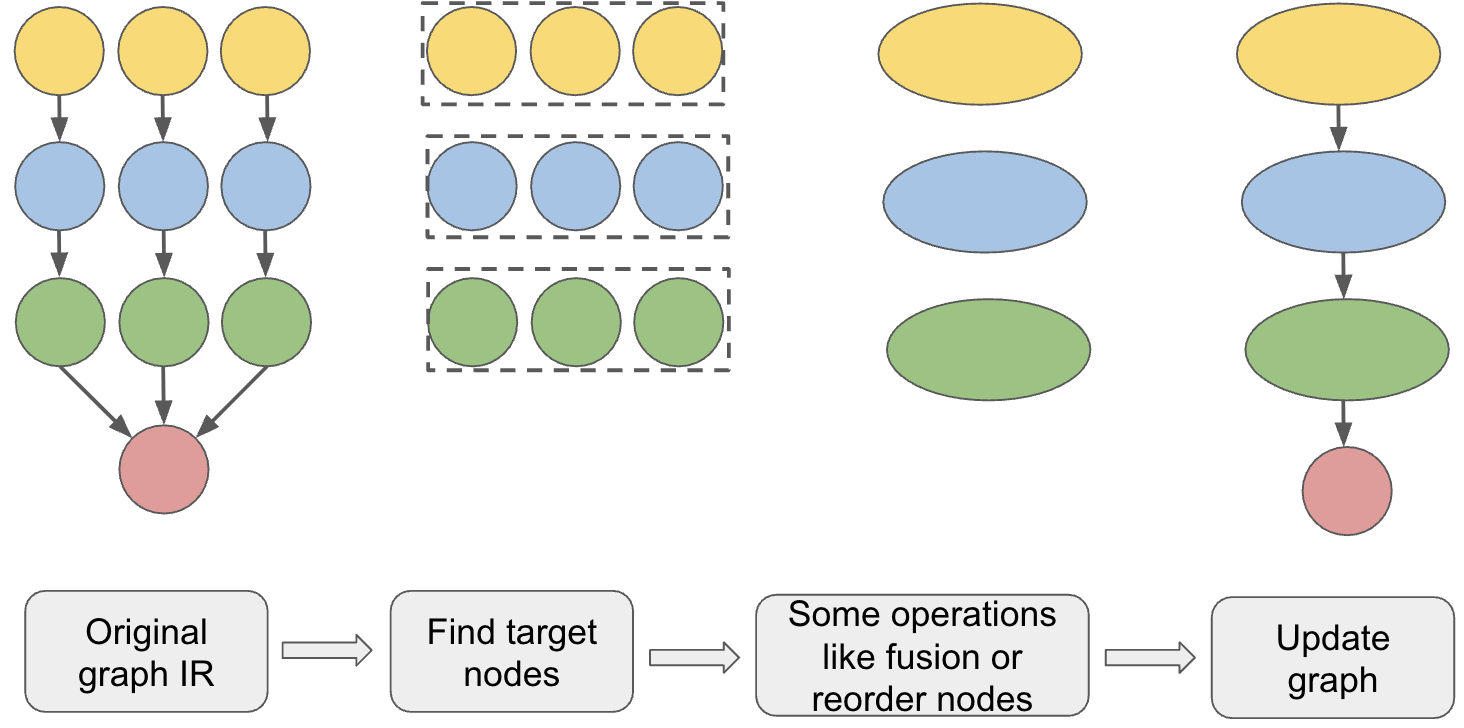}
    \caption{High-level diagram of model transformation.}
    \label{fig:procedure}
\end{figure}

The high-level procedure for conducting model transformation is illustrated in Fig.~\ref{fig:procedure}.
The process begins by identifying target patterns in the compiled graph based on predefined transformation rules.
We perform operator fusion by first locating independent groups of relevant nodes, for instance, the yellow, blue, and green nodes, each representing a distinct set that can be fused horizontally.
Next, the original nodes or subgraphs are substituted with their optimized counterparts, which may involve transformations such as operator fusion or reordering.
In this case, each color-coded group is fused into a single composite node.
Finally, the graph is recompiled to produce the optimized model.

Our model transformations are guided by heuristic rules that can be abstracted and distilled into a set of atomic principles that encompass strategies, such as horizontal and vertical optimizations.
These transformations enable structural changes to reduce training and inference latency, as well as quantization-based optimizations that lower memory usage by applying lower-precision representations to selected nodes while preserving model quality.
% These transformations enable not only structural transformations aimed at lowering training and inference latency, but only quantization optimizations to reduce memory usage while maintaining model functionality by targeting specific nodes for lower-precision representations.
% By employing lower-precision representations, quantization reduces memory usage while maintaining model functionality.
We now illustrate these concepts to highlight the capabilities of \algns.

\begin{figure}[!t]
    \centering
    \includegraphics[width=0.8\linewidth]{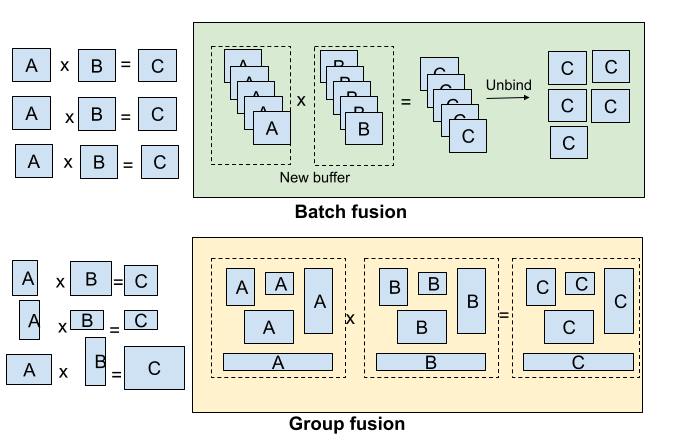}
    \caption{Horizontal transformation: batch vs group fusion.}
    \label{fig:horizontal-fusion}
\end{figure}

\textbf{1) Horizontal Optimization:} This category includes batch fusion and group fusion.
Batch fusion involves stacking tensors without data dependency into a contiguous buffer, which requires all input tensors to have identical shapes.
In contrast, group fusion allows multiple no-data dependency instances to be executed within a single kernel launch without altering the individual input tensor layouts, though it requires customized kernels.
As illustrated in Fig.~\ref{fig:horizontal-fusion}, in batch fusion, when repeatedly computing $A \times B$, identical tensors $A$ and $B$ can be grouped into two larger tensors. A single multiplication yields the grouped result $C$, which is subsequently unbound to recover the original independent outputs.
Group fusion follows a similar principle but relaxes the requirement that tensors share the same size. By fusing tensors into a larger tensor, operations can be executed via a single kernel rather than multiple small kernel launches, thereby reducing launch overhead.
The performance benefits of such fusions, however, depend on several factors, including tensor sizes, the number of tensors being grouped, and the underlying hardware architecture.

\textbf{2) Vertical Optimization:} Vertical fusion focuses on optimizing consecutive operations, many of which are already effectively handled by the PT2 kernel fusion framework.
However, additional opportunities remain, particularly in the case of split–concat patterns, where redundant nodes can be eliminated or restructured to improve the efficiency of the model graph.
For instance, consider the example in Listing~\ref{lst:code} in Appendix~\ref{sec:graph-representative}.
The input tensor undergoes two successive split operations. However, upon closer inspections, the program can be equivalently rewritten to eliminate split-concat operators as shown in Listing~\ref{lst:optimized}.
This optimization can be achieved through vertical fusion at the graph IR, without requiring any modifications to the original model code.
\begin{lstlisting}[
    language=Python, 
    caption={Optimized split-concat example.}, 
    label=lst:optimized, 
    frame=single, 
    basicstyle=\ttfamily\small, 
    columns=fullflexible,
    breaklines=true
]
def fn(x):
    return x

input = torch.randn(2, 6)
output = torch.compile(fn)(input)
\end{lstlisting}

% \begin{figure}[!t]
%     \centering
%     \includegraphics[width=0.85\linewidth]{pictures/example.png}
%     \caption{Chained model transformations}
%     \label{fig:vertical-fusion}
% \end{figure}

\begin{figure}[!t]
    \centering
    \includegraphics[width=0.8\linewidth]{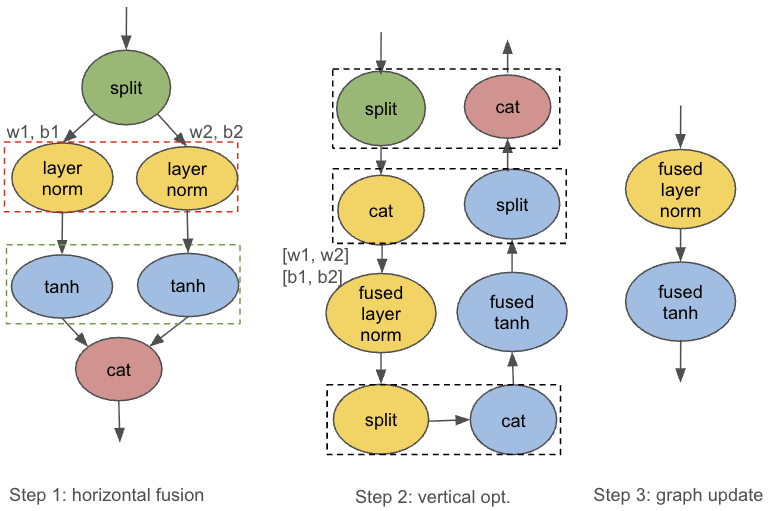}
    \caption{Chained model transformations}
    \label{fig:vertical-fusion}
\end{figure}

We present an example in Fig.~\ref{fig:vertical-fusion}, where both horizontal and vertical fusions are applied as part of the model transformation workflow.
The process begins with horizontal fusion of the layer normalization operations originating from the same split node, followed by horizontal fusion of the subsequent tanh operations. Finally, vertical fusion is applied to eliminate redundant split–concat patterns.
Together, these transformations streamline the model structure, enhancing efficiency and reducing computational overhead.
Appendix~\ref{sec:cases} presents several representative use cases of our model transformation framework.

\subsection{\algns's Graph and Rule Representations} \label{subsec:OptimusRep}
In this section, we present how \alg models computational graphs and defines transformation rules.

\textbf{1) Computational Graph Representations:} After being compiled by PT2, a PyTorch program is represented as a directed acyclic graph (DAG).
In this graph, each node corresponds to an operator (e.g., addition, multiplication), while edges denote the flow of data (tensors) between these operations.
The graph is constructed dynamically during the forward pass and subsequently leveraged to compute gradients efficiently during backpropagation.
As discussed earlier, the specific form of the graph depends on the IR layer within the PT2 stack.
In our analysis, we consider both torch operators and aten operators.

\textbf{2) Transformation Rule Representations:} Each transformation rule specifies that a given subgraph pattern (the source pattern) can be rewritten into another subgraph pattern (the target pattern). The input tensors to the source pattern are preserved as the input tensors to the target pattern within the graph, and each output tensor in the source pattern is mapped directly to a corresponding output tensor in the target pattern. To ensure semantic correctness, every transformation rule must guarantee that the set of inputs and outputs remains consistent.
We use symbolic expressions to represent both source and target patterns. For example, in Fig.~\ref{fig:split-cat} in Appendix~\ref{sec:graph-representative}, we define the transformation rule $r$ below to optimize the graph:
% \vspace{-0.5in}
 \[
\begin{aligned}
r\equiv x\big(&\text{split}[1](\text{item}_0,(\text{item}_1( \\
&\text{split}_2[1](\text{item}_2, \text{item}_3))))(\text{cat}[1](\text{y})))\rightarrow x(y),
\end{aligned}
 \]
% \vspace{-0.in}
where the left-hand side of the arrow denotes the source pattern, while the right-hand side represents the target pattern.
Node dependencies are expressed using parentheses, and we use the notation $[1]$ to indicate that both the split and concatenation operations are performed along dimension 1.
Each node is also annotated with metadata, such as tensor shapes, split sizes, and other relevant attributes.
After pattern replacement, the split and concat operations are eliminated, resulting in a more efficient program.

{\color{black}{Our framework is engineered for scalability and long-term maintainability through a modular, decoupled pattern-matching design. To ensure operational reliability, transformations utilize a conflict-free execution logic, where rules are applied sequentially with a guaranteed "no-op" fallback mechanism. Because these patterns are architected to be functionally independent, the underlying library scales fluidly without compounding system runtime complexity. This structural independence turns performance optimization into a modular plug-in architecture. New deep learning motifs can be supported simply by registering a new pattern definition through a streamlined interface, significantly reducing engineering overhead as production models continue to evolve.}}

\section{The \alg Algorithm}
In this section, we present our proposed graph search algorithm (Section~\ref{subsec:alg}), with a particular focus on how subgraph patterns are efficiently identified using a greedy-based approach. 

\begin{algorithm}[t!]
\DontPrintSemicolon
\SetNoFillComment
% \SetVline % Enables clean vertical guide lines for nested loops
\caption{Subgraph search to be transformed.}
\label{alg:node_search}

\textbf{Input:} PT2 compiled model graph $G$ and given rule $r$. \\
\textbf{Output:} subgraphs to be transformed.
\vspace{0.4em}

\SetKwFunction{FHorizontal}{HorizontalOpt}
\SetKwFunction{FIsMatched}{IsNodeMatched}
\SetKwFunction{FVertical}{VerticalOpt}

% --- HorizontalOpt ---
\SetKwProg{Fn}{Function}{}{}%
\Fn{\FHorizontal{$r$, $G$}\label{line:horizontal_opt}}{
    Extract target node $t$ and its metadata\;\label{line:meta_data}
    Perform topological sort on $G$\;\label{line:sort}
    $candidates \gets \{\}$\;
    \For{$n \in \text{reversed}(G.\text{nodes})$}{\label{line:reverse_loop}
        Run BFS from $n$ with maximum search depth \textit{max\_depth} to find matched node $t$\;\label{line:bfs}
        Retrieve node key from BFS \tcp*{key encodes metadata such as node shape}\label{line:key}
        Append $n$ to $candidates[\text{key}]$\;\label{line:append}
    }
    \For{each $(\text{key}, \text{nodes}) \in candidates$}{\label{line:condiates_loop}
        \If{$|\text{nodes}| < \textit{min\_fuse\_set\_size}$}{\label{line:skip}
            \textbf{continue}\;
        }
        Identify independent subsets within \textit{nodes}\;\label{line:subset}
    }
    \Return $candidates$\;\label{line:horizontal_return}
}
\vspace{.1in}

% --- IsNodeMatched ---
\Fn{\FIsMatched{$P.node$, $G.node$}\label{line:recursive}}{
    \If{$G.node$ matches $P.node$}{
        \For{$P.child$, $G.child$ \text{in} \text{zip}($P.children$, $G.children$)}{
            \If{\textnormal{not} \FIsMatched{$P.child$, $G.child$}}{
                \Return $\text{False}$\;
            }
            \Return $\text{True}$\;
        }
    }
    \Return $\text{False}$\;
}
\vspace{.1in}

% --- VerticalOpt ---
\Fn{\FVertical{$r$, $G$}\label{line:vertical_opt}}{
    Extract target node $t$, source pattern $P$\;\label{line:vertical_meta}
    $candidates \gets []$\;
    \For{$n$ \text{in} $G.nodes$}{\label{line:node_loop}
        \If{\FIsMatched{$t$, $n$}}{
            Add $n$ to $candidates$\;\label{line:add}
        }
    }
    \Return $candidates$\;\label{line:vertical_return}
}
\end{algorithm}

\subsection{Efficient Graph Search Algorithm Design} \label{subsec:alg}

It is well known that graph search is an NP-hard problem in general~\cite{garey1976complexity}.
To improve efficiency, our algorithm adopts a greedy strategy.
We present our proposed search algorithm in Algorithm~\ref{alg:node_search}.
Our algorithm integrates two complementary pattern search functions: horizontal optimization (Line~\ref{line:horizontal_opt}) and vertical optimization (Line~\ref{line:vertical_opt}), both of which take as input a transformation rule $r$ and a compiled graph $G$.

For horizontal optimization, we begin by extracting the target node $t$ to be searched, along with the metadata parameters from $r$. These parameters define the minimum and maximum search depths for BFS.
A smaller \textit{min\_depth} generally yields limited performance gains, while a larger \textit{max\_depth} increases search coverage at the cost of longer compile time.
Thus, they jointly serve as a trade-off between performance benefit and compile overhead.
Before starting the search, we perform a topological sort on $G$ to ensure proper graph ordering (Line~\ref{line:sort}).
The algorithm then iterates through the nodes of $G$ in reverse order (Line~\ref{line:reverse_loop}).

For each node $n$, we execute BFS up to depth \textit{max\_depth} to determine whether it matches the target node $t$ (Line~\ref{line:bfs}).
If a match is found, we extract its key, which is an encoding of metadata such as tensor shape (Line~\ref{line:key}), to guarantee that all candidate nodes for fusion share consistent attributes.
The node is then added to the candidate list associated with its key (Line~\ref{line:append}).
Once all candidates are collected, we iterate over each key-nodes pair (Line~\ref{line:condiates_loop}).
Candidate sets with fewer nodes than \textit{min\_fuse\_set\_size} are skipped (Line~\ref{line:skip}), as they are unlikely to provide meaningful performance benefits.
For sufficiently large sets, we further identify independent subsets of nodes using another BFS traversal (Line~\ref{line:subset}).
Nodes are considered independent if no data dependency exists between them.
Finally, for each such subset, we apply the corresponding graph transformation defined by the target pattern in rule $r$.
\alg is designed to group independent operators for horizontal fusion, while preserving their original order of execution.

For vertical optimization, we begin by extracting the target node $t$ and the source pattern $P$ (Line~\ref{line:vertical_meta}).
The core of the algorithm relies on a recursive matching procedure that traverses both the compiled graph $G$ and the pattern graph $P$ for comparison.
Starting from an anchor node (Line~\ref{line:node_loop}), which corresponds to the output of the pattern, we recursively explore the child nodes of both graphs, verifying alignment at each step until a complete match is found (Line~\ref{line:recursive}).
When a match is identified, the corresponding node in $G$ is added to the candidate set (Line~\ref{line:vertical_return}).
Although general graph matching is an NP-hard problem, the class of graphs typically encountered in machine learning programs is considerably simpler.
In practice, the problem is closer to tree inclusion~\cite{bille2011tree}, which can be solved with algorithms that use linear space and achieves sub-quadratic running times.
Specifically, it can be characterized as i) \textit{Space Complexity}: \(O(n_{T})\), which is linear in the size of the target pattern \(T\);
ii) \textit{Time Complexity}: \(O(\min (l_{P}n_{T},l_{P}l_{T}\log \log n_{T}+n_{T},\frac{n_{P}n_{T}}{\log n_{T}}+n_{T}\log n_{T}))\), where \(n_{P}\) and \(n_{T}\) are the number of nodes, and \(l_{P}\) and \(l_{T}\) are the number of leaves in the source pattern \(P\) and target pattern \(T\).

\begin{algorithm}[t]
\DontPrintSemicolon
\SetNoFillComment
\caption{Apply \alg graph transformation rules.}
\label{alg:customized-optimus}

\textbf{Input:} PT2 compiled model graph $G$, set of rules $R$. \\
\textbf{Output:} updated graph $G$.
\vspace{0.4em}

\For{rule $r \in R$}{\label{line:loop_r}
    \If{$r$ is horizontalOptimizaton rule}{\label{line:h_r}
        graphs $\gets$ HorizontalOpt($r$, $G$) in Alg.~\ref{alg:node_search}\;\label{line:horizontal_alg}
        Transform graphs based on rule $r$\;\label{horizontal_transform}
    }
    \Else{\label{line:v_r}
        Get $counter$ from $r$ \tcp*{repetitive time to apply $r$}\label{line:counter}
        \For{$i$ \textnormal{in} \textnormal{range}($counter$)}{\label{line:loop_c}
            graphs $\gets$ VerticalOpt($r$, $G$) in Alg.~\ref{alg:node_search}\;\label{line:vertical_alg}
            Transform graphs based on rule $r$\;\label{line:vertical_transform}
        }
    }
}
\Return $G$\;
\end{algorithm}

Next, we present our graph transformation algorithm in Algorithm~\ref{alg:customized-optimus}.
The algorithm iterates over each transformation rule $r$ in sequence (Line~\ref{line:loop_r}).
For horizontal optimization rules, the procedure invokes \textsc{HorizontalOpt} to identify candidate nodes (Line~\ref{line:horizontal_alg}), followed by applying the corresponding graph transformations to these candidates (Line~\ref{horizontal_transform}).
For vertical optimization rules, the algorithm first retrieves the metadata counter from $r$ (Line~\ref{line:counter}), which specifies the number of times the rule should be applied iteratively.
During each iteration, \textsc{VerticalOpt} is invoked to identify candidates (Line~\ref{line:vertical_alg}), after which the appropriate graph transformation is applied according to the rule $r$ (Line~\ref{line:vertical_transform}).
{\color{black} When parameters are not explicitly defined by the user, the framework will use predefined default configurations (e.g., min\_fuse\_set\_size = 5, max\_depth = 5).}

We control the combinatorial explosion by making pattern selection architecture-aware.
We first identify anchor operators based on the model structure and profiling insights, then search only their independent operators or immediate neighbors to check whether predefined patterns apply.
For example, starting from an anchor such as {\tt torch.split}, we examine its neighboring operators to capture dependencies; if the local subgraph matches a predefined pattern, it is added as a candidate for graph transformation.
This localized and anchor-driven search effectively prunes the search space and guides pattern selection.

% \begin{table*}[ht]
%     \centering
%     \includegraphics[width=.8\linewidth]{pictures/internal_model.png}
%     \caption{\alg Graph Transformation Patterns with QPS Improvement Across Industry-Scale Recommendation Models.}
%     \label{table:internal_model}
% \end{table*}

\begin{table*}[htbp]

\centering
\small
\caption{Graph transformation patterns with QPS improvement across industry-scale recommendation models.}
\label{table:internal_model}
\begin{tabular}{|l|M{2.1cm}|M{5cm}|M{3cm}|M{3.1cm}|M{.6cm}|}
\hline
\multirow{4}{*}{\textbf{}} & \multicolumn{2}{c|}{\textbf{Torch-IR Graph Transformation Patterns}} & \multicolumn{2}{c|}{\textbf{Aten-IR Graph Transformation Patterns}} & \multirow{4}{1.4cm}{\textbf{QPS \\Gain}} \\ \cline{2-5}
 & \textbf{Horizontal Optimization} & \textbf{Vertical Optimization} & \textbf{Horizontal Optimization} & \textbf{Vertical Optimization} &  \\ \hline
 & 
 \texttt{batch\_linear\_lhs} \newline 
 \texttt{batch\_linear} \newline 
 \texttt{batch\_layernorm} \newline 
 \texttt{batch\_tanh} \newline 
 \texttt{batch\_sigmoid} \newline 
 \texttt{batch\_relu} \newline 
 \texttt{batch\_detach} \newline 
 \texttt{batch\_clamp} \newline 
 \texttt{batch\_nan\_to\_num} 
 & 
 \texttt{normalization\_pass} \newline 
 \texttt{remove\_split\_with\_size\_one\_pass} \newline 
 \texttt{merge\_getitem\_cat\_pass} \newline 
 \texttt{merge\_splits\_pass} \newline 
 \texttt{mutate\_cat\_pass} \newline 
 \texttt{split\_cat\_pass} \newline 
 \texttt{unbind\_stack\_pass} \newline 
 \texttt{split\_cat\_to\_slices\_pass} \newline 
 \texttt{unbind\_cat\_to\_view\_pass} \newline 
 \texttt{split\_stack\_to\_cats\_pass} \newline 
 \texttt{unbind\_stack\_to\_slices\_pass} \newline 
 \texttt{move\_reshape\_out\_of\_split\_stack\_pass} \newline 
 \texttt{einsum\_to\_pointwise\_pass} 
 & 
 \texttt{batch\_linear\_post\_grad} \newline 
 \texttt{batch\_aten\_tanh} \newline 
 \texttt{batch\_aten\_sigmoid} \newline 
 \texttt{batch\_aten\_relu} \newline 
 \texttt{batch\_aten\_sub} \newline 
 \texttt{batch\_aten\_div} \newline 
 \texttt{batch\_aten\_mul} 
 & 
 \texttt{normalization\_aten\_pass} \newline 
 \texttt{unbind\_stack\_aten\_pass} \newline 
 \texttt{split\_cat\_aten\_pass} \newline 
 \texttt{select\_cat\_aten\_pass} \newline 
 \texttt{pad\_aten\_mm\_pass} \newline 
 \texttt{decompose\_mm\_pass} 
 & \\ \hline
\textbf{Model Type A} & 3\%   & 37\%  & 0\%  & 23\% & 63\%   \\ \hline
\textbf{Model Type B} & 2\%   & 12\%  & 1\%  & 10\% & 25\%   \\ \hline
\textbf{Model Type C} & 2.5\% & 10\%  & 0\%  & 14\% & 26.5\% \\ \hline
\textbf{Model Type D} & 0\%   & 12\%  & 2\%  & 4\%  & 18\%   \\ \hline
\textbf{Model Type E} & 2\%   & 11\%  & 0\%  & 0\%  & 13\%   \\ \hline
\end{tabular}

\end{table*}

\section{Performance Evaluation}
In this section, we evaluate the \alg on our internal production recommendation models. These models are variants of the DLRM architecture which is the standard for recommendation and ranking workloads~\cite{naumov2019deep}.
While proprietary constraints preclude the disclosure of specific weights and internal datasets, we have addressed this by evaluating \alg across a suite of production-scale architectural archetypes as shown in Table~\ref{table:compile_time}. 
These models are selected because they mirror the specific computational motifs and memory-bottleneck characteristics of our production environments, providing a verifiable basis for our performance claims.
{\color{black} {While \alg is IR-agnostic and supports Torch-IR, ATen-IR, and Inductor-IR via modular pattern registration, our evaluation focuses on Torch-IR and ATen-IR, since these layers provide the high-level semantic fusions most critical for recommendation models. Although Inductor-IR can be technically supported, it has not been officially launched in production models.}}

In addition, we demonstrate the effectiveness of the proposed patterns using an illustrative model developed within TorchBench~\cite{torchbench2020}, an open-source benchmark suite that includes a diverse collection of representative workloads.
Our illustrative model is open-sourced to enable reproducibility.
For industry-scale recommendation models, experiments are conducted on NVIDIA A100, H100, or B200 GPUs, with GPU counts ranging from 64 to 256.
For the TorchBench illustrative model, experiments are performed on a single NVIDIA A100 GPU with CUDA 12.4 and an Intel Xeon 8339HC CPU.
Each experiment is repeated 100 times to mitigate measurement noise, with three warm-up iterations.
All experiments use TorchInductor from the PyTorch nightly build (dated Oct. 26, 2025)
% We impose a 30-minute timeout per model, counting timeouts as failures.
% All experiments use TorchInductor from the PyTorch nightly build (dated Oct. 26, 2025), with max-autotune, freezing, and CUDA graphs optimizations enabled.

\begin{table}[htbp]
\centering
\footnotesize % Reduces font size to match typical column-width tables
\setlength{\tabcolsep}{2pt} % Tighter padding between columns
\caption{Compile time and QPS trade-off.}
\label{table:compile_time}
\begin{tabular}{llccc}
\toprule
\textbf{Model Type} & \textbf{Arch Type} & \textbf{Compile Budget} & \textbf{QPS Gain} & \textbf{Hardware} \\ 
\midrule
\multirow{2}{*}{Model A} & \multirow{2}{*}{DHEN-like~\cite{zhang2022dhen}}        & 10 min & 60\% & H100 \\
                         &                                    & 5 min  & 35\% & H100 \\ 
\midrule
Model B                  & DHEN-like~\cite{zhang2022dhen}                          & --     & --   & --   \\ 
\midrule
Model C                  & Wukong-like~\cite{wukong2024}                        & --     & --   & --   \\ 
\midrule
Model D                  & COFFEE-like~\cite{roychowdhury2026coffee}                        & --     & --   & --   \\ 
\midrule
\multirow{2}{*}{Model E} & \multirow{2}{*}{Interformer-like~\cite{zeng2025interformer}} & 10 min & 10\% & B200 \\
                         &                                    & 5 min  & 4\%  & B200 \\ 
\bottomrule
\end{tabular}
\end{table}

\begin{figure*}[h!]
\begin{minipage}{0.32\linewidth}
%\centering  % redundant
\includegraphics[width=\textwidth]{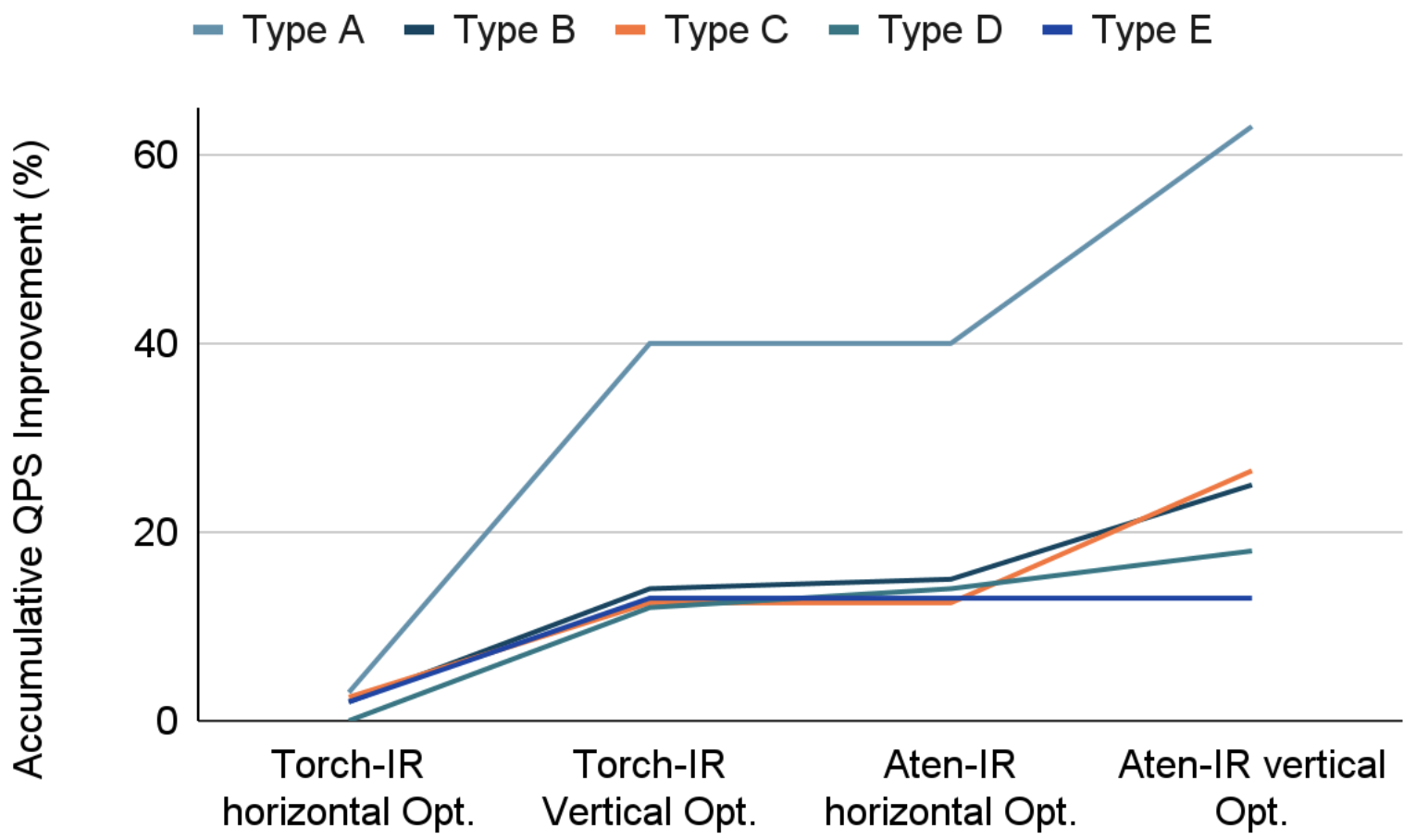}
\caption{QPS gain with \alg graph transformation.}
\label{fig:internal_line_figure}
\end{minipage}%
\hfill% not: "\hspace{0.5cm}"
\begin{minipage}{0.32\linewidth}
%\centering  % redundant
\includegraphics[width=\textwidth]{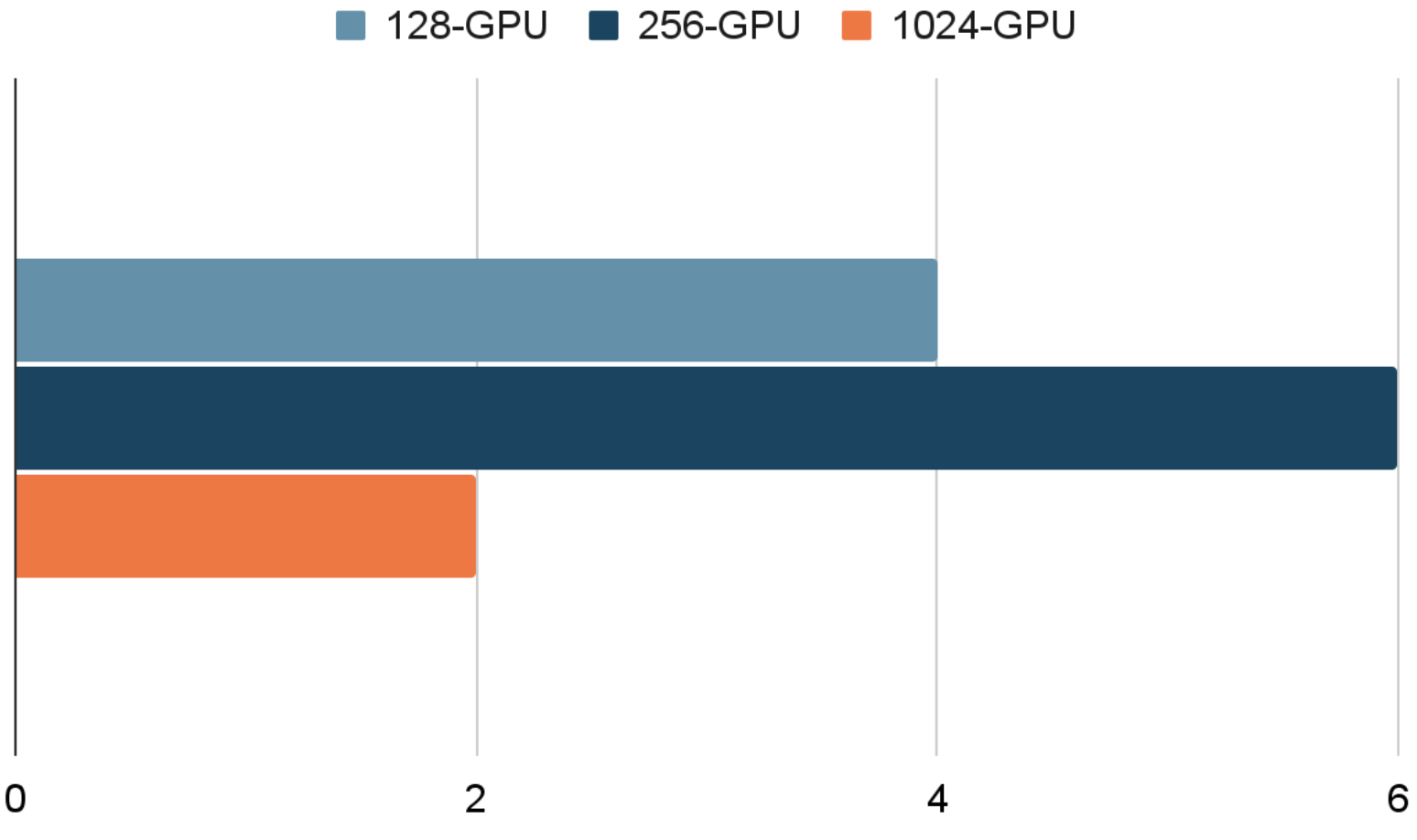}
\caption{Peak memory reduction in percentage on Foundation Model.}
\label{fig:memory_save}
\end{minipage}%
\hfill% not: "\hspace{0.5cm}"
\begin{minipage}{0.32\linewidth}
%\centering  % redundant
\includegraphics[width=\textwidth]{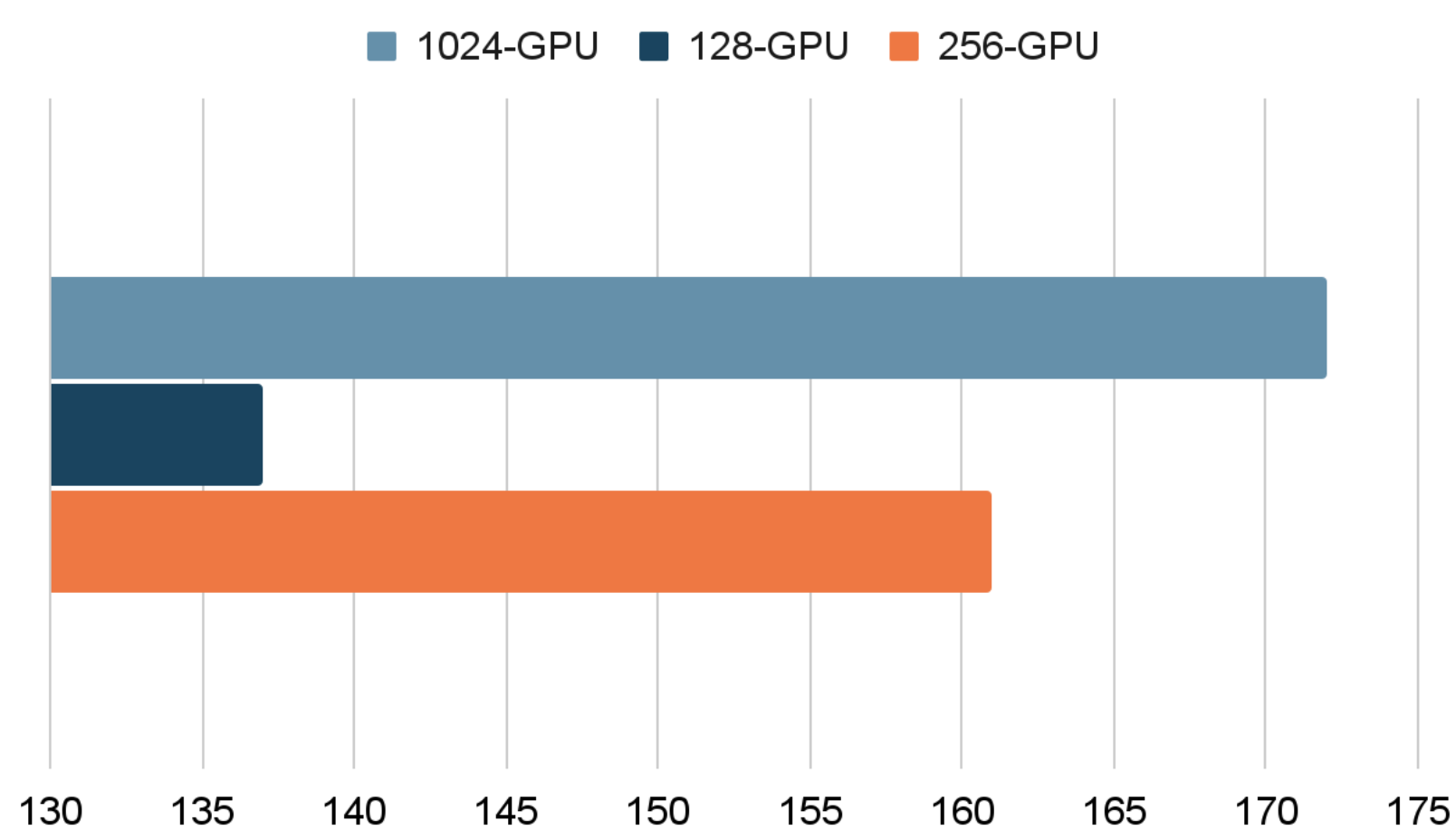}
\caption{Compile time overhead in second on foundation model.}
\label{fig:compile_time}
\end{minipage}
\end{figure*}

\subsection{Real-World Models Evaluation}
We begin by evaluating industry-scale recommendation models.
As summarized in Table~\ref{table:internal_model}, we enumerate all transformation patterns that contribute to the improvement of QPS, categorizing them according to the level at which the model transformation is applied (Torch-IR or Aten-IR) and by the type of optimization, that is, horizontal or vertical.
We focus on the top five high-throughput recommendation models and report the QPS gain achieved by each optimization category (see Fig.~\ref{fig:internal_line_figure}).
From the results, we observe that the majority of the QPS gains originate from vertical split-concat optimizations, which effectively eliminate redundant split operations.
This reduction substantially decreases the number of generated and launched Triton kernels, leading to notable performance gains.
It is worth noting that batch fusion introduces additional \textit{stack} and \textit{unbind} operators, which can incur latency overhead.
To fully realize the benefits of horizontal fusion, a sufficient number of compatible operators must be available for batching.
When this condition is met, the reduced kernel launch overhead can outweigh the additional cost introduced by the auxiliary operations.
% \begin{figure}
%     \centering
%     \includegraphics[width=1\linewidth]{pictures/internal_model_line_figure.png}
%     \caption{QPS Improvement with \alg Graph Transformation.}
%     \label{fig:internal_line_figure}
% \end{figure}

{\color{black}To evaluate the feasibility of large-scale industrial deployment, we characterize the trade-off between compilation overhead and training efficiency with distinct architectural types and hardware configurations (Table~\ref{table:compile_time}). Our results show that while expanding the graph-transformation search space increases compile time, it yields substantial, non-linear dividends in runtime throughput. For the computationally dense DHEN-like architecture (Model A) on NVIDIA H100 GPUs, extending the compile budget from 5 to 10 minutes unlocks an additional 25\% absolute increase in QPS gain (improving from 35\% to 60\%), demonstrating the discovery of highly profitable global graph-fusion and multi-linear grouping patterns. Conversely, for the memory-bound Interformer-like architecture (Model E) on NVIDIA B200 hardware, doubling the budget yields a more modest performance increment (from 4\% to 10\% QPS gain). This variation underscores that while extended compilation universally maximizes runtime throughput, the magnitude of the benefit is inherently tied to the structural motifs of the model and the underlying hardware execution paradigm.}
% allowing operators to strategically balance compilation time against production throughput SLAs.

\begin{table}[t!]
\centering
\footnotesize % Downsizes text to a compact, publication-grade footprint
\setlength{\tabcolsep}{3pt} % Compresses empty space between columns
\renewcommand{\arraystretch}{1.1} % Maintains professional line height
\caption{Performance comparison of opt. techniques.}
\label{tab:optimization_speedups}

\begin{tabular}{llcc}
\toprule
\textbf{Optimization} & \textbf{Patterns} & \textbf{Inf. Speedup} & \textbf{Train Speedup} \\
\midrule
\multirow{4}{*}{\textbf{Vertical}} 
 & merge\_splits\_pass         & \multirow{4}{*}{48.847x} & \multirow{4}{*}{1.253x} \\ 
 & split\_cat\_pass           &                          &                         \\ 
 & unbind\_stack\_pass        &                          &                         \\ 
 & unbind\_cat\_to\_view\_pass &                          &                         \\ 
\midrule
\multirow{2}{*}{\textbf{Horizontal}} 
 & batch\_linear              & \multirow{2}{*}{15.045x} & \multirow{2}{*}{0.826x} \\ 
 & batch\_layernorm           &                          &                         \\ 
\midrule
\textbf{Combined} 
 & All above patterns         & 364.035x                 & 11.160x                 \\
\bottomrule
\end{tabular}
\end{table}

We next evaluate how the \textit{activation\_quantization\_aten\_pass} contributes to peak memory reduction in the large recommendation foundation models executed on NVIDIA H100 GPUs.
Experiments are conducted across 128-, 256-, and 1024-GPU configurations. The 128-GPU and 256-GPU runs correspond to scaled-down versions of the full model.
Overall, we observe a 2–6\% reduction in peak memory usage (see Fig.~\ref{fig:memory_save}).

As expected, applying this pattern increases compilation time, and the corresponding results are presented in Fig.~\ref{fig:compile_time}.
In this evaluation, we use the default configuration for activation quantization. However, \alg is designed as a general and flexible framework that allows users to tune multiple parameters to find the optimal trade-off for their specific models.
It is important to emphasize that no single configuration can simultaneously optimize QPS, memory savings, and numerical equivalence (NE). Achieving the best overall performance requires carefully balancing these objectives based on system priorities.
Specifically, aggressive activation quantization, for example, lowering the \textit{size\_in\_mb} threshold, including input tensors by setting \textit{exclude\_primals=False}, or expanding the \textit{allowed\_dtypes} list, can lead to greater memory savings but increases the risk of NE degradation.
% Intuitively, quantizing from FP16 to FP8 generally presents a lower NE risk than quantizing from FP32 to FP8.
% Similarly, quantizing more nodes increases NE risk compared to quantizing fewer nodes, assuming all other configurations are held constant.
Moreover, disabling scaling (by setting \textit{use\_scaling=False}) can yield additional memory savings by avoiding storage of auxiliary scale tensors, but this configuration may introduce NaN issues if the model inputs overflow in the FP8 date type.
If minimizing NE risk is a higher priority than maximizing memory savings, it is advisable to enable scaling quantization.
% Finally, for QPS improvement, two complementary strategies can be considered:
% (1) quantizing a larger subset of nodes to lower-precision types (e.g., BF16 or FP8), or
% (2) selectively quantizing fewer nodes while retaining higher-precision types (e.g., FP32) to preserve NE stability.
% The primary source of QPS gains arises from the reduced latency of clone kernels when using FP8 compared to the original higher-precision data types.
\begin{figure*}[t!] % Adding the * forces the figure to cross both columns
    \centering
    % Left Subfigure (takes up 48% of the total page width)
    \begin{subfigure}[b]{0.48\textwidth}
        \centering
        \includegraphics[width=\linewidth]{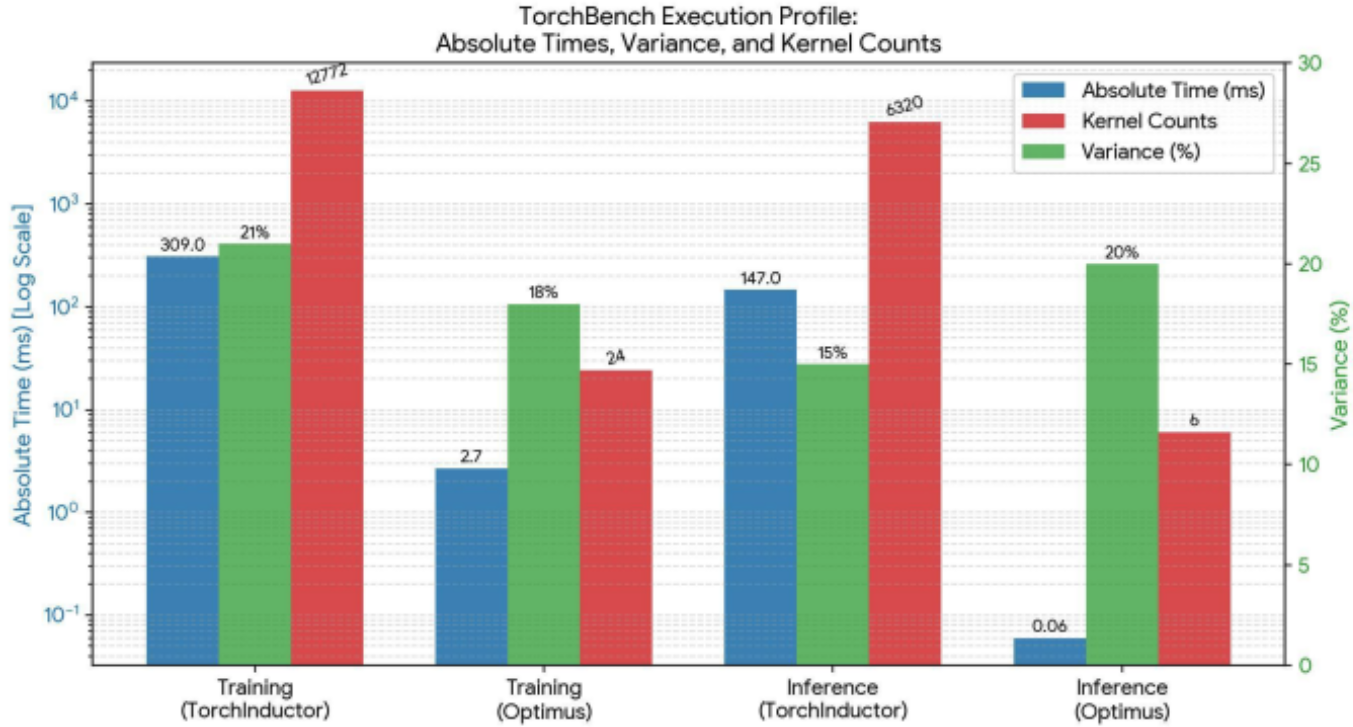}
        \caption{Absolute execution profile.}
        \label{fig:torchbench_profile}
    \end{subfigure}
    \hfill % Pushes the figures smoothly to opposite margins
    % Right Subfigure (takes up 48% of the total page width)
    \begin{subfigure}[b]{0.48\textwidth}
        \centering
        \includegraphics[width=\linewidth]{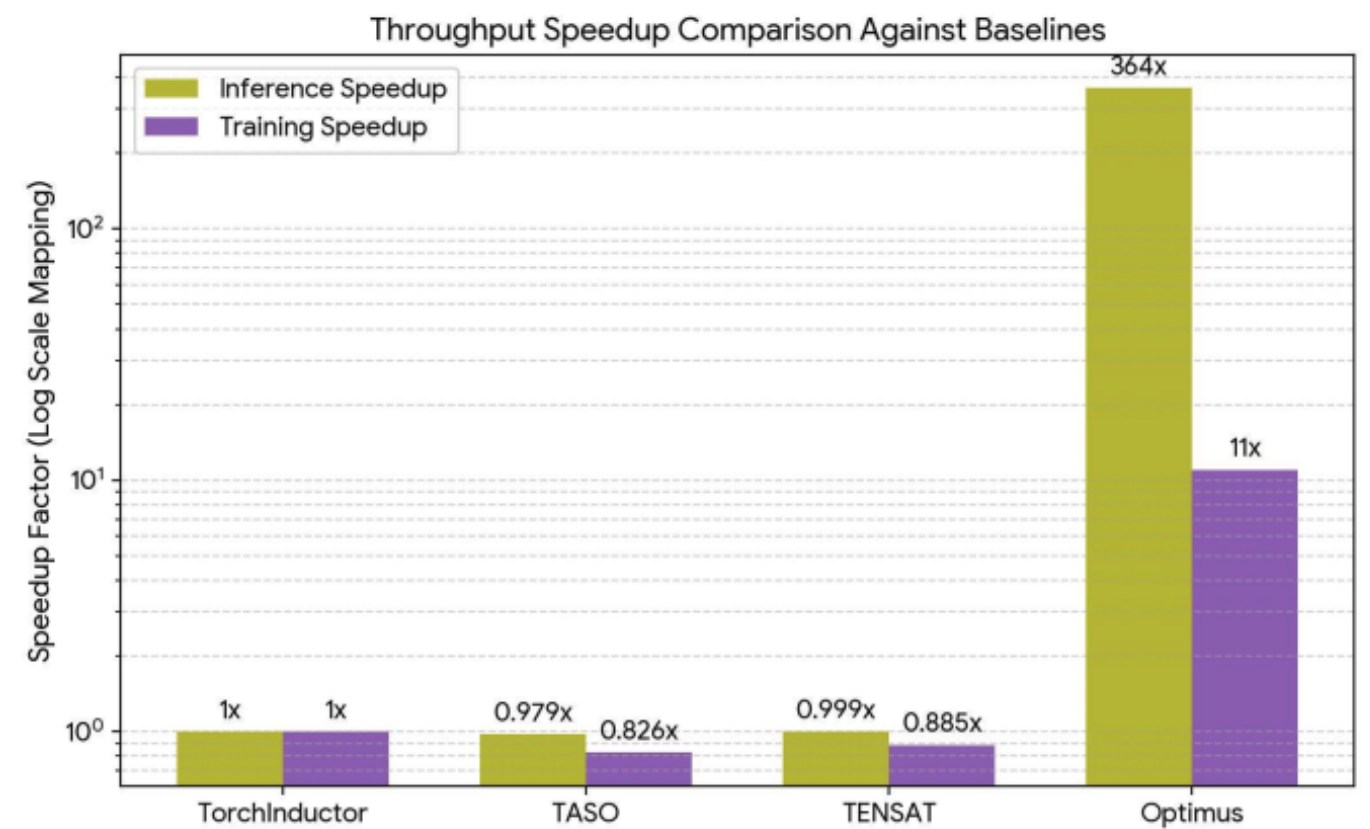}
        \caption{Baseline speedup comparison.}
        \label{fig:throughput_comparison}
    \end{subfigure}
    
    \caption{TorchBench evaluation results: (a) Execution profile highlighting absolute times, variance, and kernel counts; (b) Throughput speedup comparisons against baselines across inference and training phases.}
    \label{fig:evaluation_results}
\end{figure*}

\subsection{TorchBench Illustrative Model Evaluation}

We added an illustrative model to TorchBench~\footnote{Available at https://github.com/pytorch/pytorch/blob/main/benchmarks/dynamo/optimus.py.} to demonstrate the effectiveness of \alg using a selective set of patterns that we identified as strong candidates for latency reduction.
Specifically, we designed patterns within the model to emulate real-world scenarios observed in our recommendation models, while simplifying the model architectures and reducing input scales for ease of evaluation. The tested tensor has a shape of $64\times 64$ with data type torch.bfloat16.
We conducted experiments for both inference and training to measure performance gains compared to the TorchInductor baseline, where no \alg patterns were applied.
As shown in Table~\ref{tab:optimization_speedups}, both vertical and horizontal optimization patterns contribute to reducing training and inference latency by effectively decreasing the number of launched kernels.
However, in our example, vertical optimization patterns achieve greater improvements than horizontal ones in training.
This is because the input tensor of shape $64\times 64$ is relatively small, limiting the benefits of batch fusion, as the number of tensors available for batching is not large enough to fully exploit horizontal fusion.

{\color{black}To evaluate the efficiency of \alg, we analyzed the GPU execution traces for the TorchInductor and Optimus backends (Fig.~\ref{fig:torchbench_profile}). The profile shows that \alg significantly compresses the kernel footprint—reducing training kernels from 12,772 to 24, and inference kernels from 6,320 to 6. This massive consolidation eliminates kernel launch overhead, driving the observed latency reductions.}
{\color{black} Furthermore, our evaluation indicates that TASO~\cite{jia2019taso} and TENSAT~\cite{yang2021equality} lack the necessary graph optimization passes required to transform this illustrative model. Consequently, as demonstrated in Fig.~\ref{fig:throughput_comparison}, neither framework applies any structural optimizations to the graph, yielding zero performance improvement. }
% As demonstrated in Fig.~\ref{fig:throughput_comparison}, neither framework yields any inference performance or speedup improvements on the TorchBench workloads.}

Moreover, the order in which the optimization patterns are applied plays a crucial role since the horizontal fusion stacks multiple tensors together before subsequently unbinding them, which can introduce nontrivial overhead when the number of tensors available for batching is limited.
In our implementation, horizontal optimizations are applied first, introducing additional \textit{stack} and \textit{unbind} nodes.
Subsequently applying vertical optimizations eliminates these redundant nodes, resulting in a more compact graph and better performance.
% It is also worth noting that horizontal fusion does not come without cost.
% This optimization frist stacks multiple tensors together before subsequently unbinding them, which can introduce nontrivial overhead when the number of tensors available for batching is limited.
As observed in our results, the performance improvement achieved by combining horizontal and vertical optimizations is significantly greater than that of either technique alone.
This synergistic effect arises because vertical optimization patterns effectively eliminate the redundant nodes and overhead introduced by horizontal fusion, thereby removing unnecessary split and concatenation operations.
This example illustrates that both the selection of optimization patterns and their application order are highly model-dependent.
Therefore, automating the pattern selection and order scheduling based on model characteristics represents a promising direction for future work.

Another observation is that the training and inference speedups can differ substantially because they stress the system in different ways and benefit from different compiler optimizations.
Inference primarily runs the forward pass, where aggressive operator fusion and kernel consolidation can dramatically reduce launch overhead and intermediate memory traffic, yielding very large latency improvements.
Training, by contrast, includes the backward pass and sometimes optimizer work, which has a different operator mix, more gradient-related memory movement, and often fewer or different fusion opportunities, as a result, forward pass gains are diluted by backward pass cost.
Additionally, training graphs are more likely to experience graph breaks, extra synchronization, or less favorable caching or specialization behavior, all of which can limit end-to-end speedups compared to the inference path.

\section{Related work}
  
With the rapid advancement and widespread deployment of deep learning, reducing the training and inference costs of DNNs has become increasingly critical as models continue to grow in scale and complexity.  
A prominent line of research focuses on optimizing DNN computation graphs via graph substitutions.  
TASO~\cite{jia2019taso}, the first computation graph optimizer for DNNs, employs a cost-based backtracking search to automatically generate optimized graph substitutions.  
Building on this, Fang et al.~\cite{fang2020optimizing} introduce a more efficient sampling-based search algorithm that prunes redundant substitutions, thereby improving search efficiency.  
However, sampling-based approaches risk missing globally optimal substitutions.  
To overcome this limitation, TENSAT~\cite{yang2021equality} leverages equality saturation to apply all possible substitutions in parallel.  
 
Another related line of research is \textit{superoptimization}, which seeks to automatically discover correct and optimal programs under a given cost model.  
Most prior efforts~\cite{massalin1987superoptimizer,bansal2006automatic,schkufza2013stochastic,schkufza2014stochastic,phothilimthana2016scaling,phothilimthana2016greenthumb,sasnauskas2017souper} focus on short sequences of low-level instructions and do not rely on explicit rewrite rules.  
Denali~\cite{joshi2002denali} was the first to demonstrate how e-graphs can be applied to program optimization by using rewrite rules in conjunction with a constraint solver, trading off performance for improved scalability, but none of them directly addresses tensor computation graphs. 
This gap was first bridged by TENSAT~\cite{yang2021equality}, which introduced equality saturation into tensor graph optimization, allowing simultaneous graph rewrite of all possibilities.

Despite these advancements, existing approaches remain constrained by inefficient graph search and substitution mechanisms.  
As a result, their applicability to large, real-world models is limited, as compile time often becomes a significant bottleneck when dealing with complex architectures.  
In contrast, our framework defines graph transformation rules at the level of atomic operators, which enhances generalizability and enables consistent optimization across diverse models.

\section{Conclusions}
We proposed \algns, a general-purpose graph transformation framework designed to efficiently rewrite subgraphs based on concise atomic rules.
The framework is designed to support transformations across all PT2 graph IRs, enabling significant reductions in execution latency, peak memory usage, and compilation time. \alg has been deployed in our internal production systems, benefiting a large number of model variants developed through ML engineers’ experimentation and exploration.
It is fully open-sourced with PT2 stack with a highly customizable design. 
% Our evaluation demonstrates that \alg achieves up to 63\% speedup, 6\% peak memory reduction, and over 400 seconds of compile time savings compared to baseline systems on our top recommendation models.

In future work, we plan to extend \alg by incorporating additional transformation patterns not only for Torch-IR and Aten-IR, but also for Inductor-IR to further improve performance.
% Owing to the flexibility of the framework, we also aim to introduce advanced features, such as CPU offloading to alleviate memory bottlenecks in large foundation models.

% \section{Acknowledgments}

% Identification of funding sources and other support, and thanks to
% individuals and groups that assisted in the research and the
% preparation of the work should be included in an acknowledgment
% section, which is placed just before the reference section in your
% document.

% This section has a special environment:
% \begin{verbatim}
%   \begin{acks}
%   ...
%   \end{acks}
% \end{verbatim}
% so that the information contained therein can be more easily collected
% during the article metadata extraction phase, and to ensure
% consistency in the spelling of the section heading.

% Authors should not prepare this section as a numbered or unnumbered {\verb|\section|}; please use the ``{\verb|acks|}'' environment.

% \clearpage
\section{Appendices}
\subsection{PyTorch 2.x Compiled Model Representative}\label{sec:graph-representative}
In PyTorch, an unlimited variety of program structures can represent the same semantics.
However, when compiled with PT2, the code is transformed into an IR comprising a limited set of operators~\cite{chillee2021operators}.
Consequently, performing graph transformations within the compiler becomes inherently more transferable across different models, particularly when pattern-matching rules are defined in terms of atomic operators.

We present a split-concat example to demonstrate how models are represented in Torch IR. The program hierarchically splits a tensor $x$, and subsequently returns a concatenated tensor composed of the resulting components. This is an oversimplified example but the high-level structure is commonly observed in recommendation workloads mostly after the embedding lookup operations. 

Within \textit{torch.compile}, we traverse the various layers in PT2 to obtain different IRs. 
For clarity, we focus exclusively on the Torch IR, as the other IRs follow a similar underlying logic but differ in their operator representations.
The Torch IR is expressed in terms of torch tensors, where nodes represent operators, and edges denote the dependency flow between them.

\begin{figure}[htbp]
    \centering
    % Left Subfigure - Scaled down to 38% of the text width
    \begin{subfigure}[b]{0.22\textwidth}
        \centering
        \includegraphics[width=\linewidth]{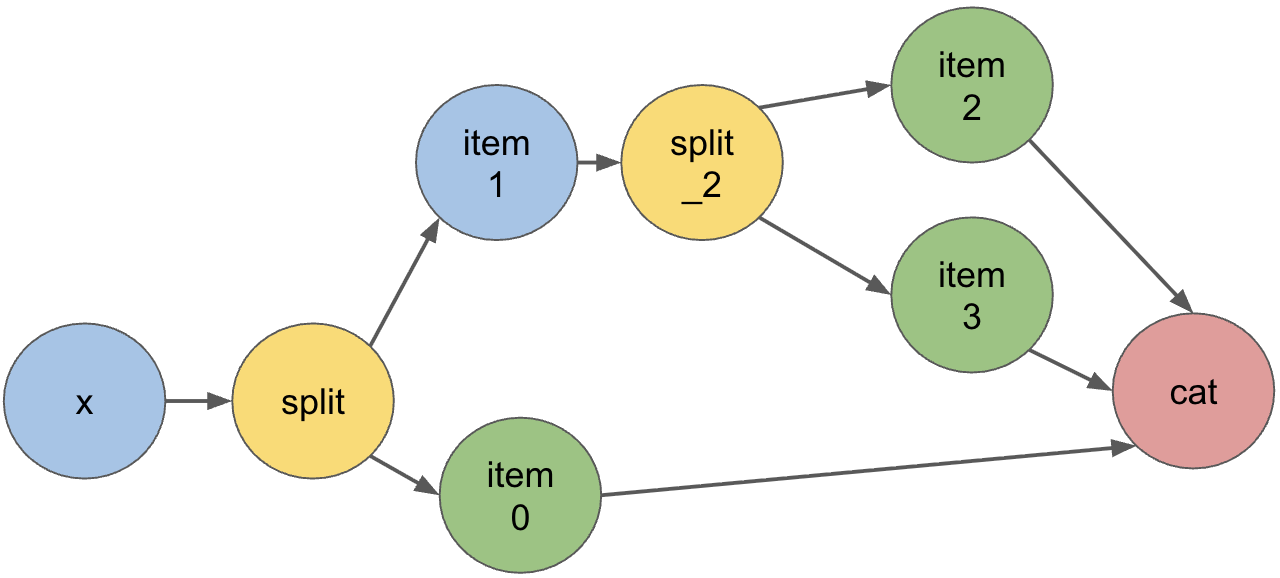}
        \caption{Split-concat transformation.}
        \label{fig:split-cat}
    \end{subfigure}
    \hspace{0.04\textwidth} % Explicit 4% horizontal spacing block between images
    % Right Subfigure - Scaled down to 38% of the text width
    \begin{subfigure}[b]{0.2\textwidth}
        \centering
        \includegraphics[width=\linewidth]{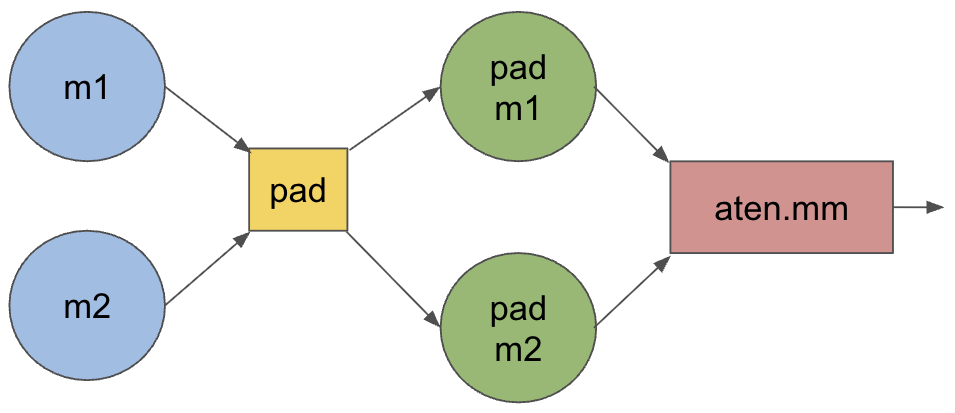}
        \caption{GEMM shape alignment padding.}
        \label{fig:pad_mm_pattern}
    \end{subfigure}
    
    % Main overarching caption (remains at standard full-page text size)
    \caption{(a) Graph transformation representation of a split-concat example; (b) Shape alignment padding optimization for GEMM operations.}
    \label{fig:graph_optimization_patterns}
\end{figure}

\begin{lstlisting}[
    language=Python, 
    caption={A simple split-concat example.}, 
    label=lst:code, 
    frame=single, 
    basicstyle=\ttfamily\small, 
    columns=fullflexible,
    breaklines=true
]
def fn(x):
    fs = torch.split(x, [2, 4], dim=1)
    item0 = fs[0]
    item1 = fs[1]
    i = [item0]
    i.extend(item1.split((2, 2), 1))
    return torch.cat(i, dim=1)

input = torch.randn(2, 6)
output = torch.compile(fn)(input)
\end{lstlisting}

\begin{lstlisting}[
    caption={Torch-IR of the split-concat example.}, 
    label=lst:torch-ir, 
    frame=single, 
    basicstyle=\ttfamily\small, 
    columns=fullflexible,
    breaklines=true
]
graph():
    %l_x_ : torch.Tensor[num_users=1] = placeholder[target=L_x_]
    %split : [num_users=3] = call_function[target=torch.functional.split](args(%l_x_, [2, 4]), kwargs={dim: 1})
    %item0 : [num_users=1] = call_function[target=operator.getitem](args=(%split, 0), kwargs={})
    %item1 : [num_users=1] = call_function[target=operator.getitem](args=(%split, 1), kwargs={})
    %split_2 : [num_users=6] = call_function[target=torch.functional.split](args=(%item1, [2, 2]), kwargs={dim: 1})
    %item2 : [num_users=1] = call_function[target=operator.getitem](args=(%split_2, 0), kwargs={})
    %item3 : [num_users=1] = call_function[target=operator.getitem](args=(%split_2, 1), kwargs={})
    %cat : [num_users=1] = call_function[target=torch.cat](arg=([%item0, %item2, %item3]), kwargs={dim: 1})
    return (cat,)
\end{lstlisting}

% To better understand the Torch-IR graph compiled by PT2.x for the example code in Listing~\ref{lst:code}.
% We give its graph representation in Fig.~\ref{fig:split-cat}.
% Basically what the example does is to split the input $x$ into three items, and further split the last item to 6 items, then all of these items concat together into one big node, which is our final return node.

% The first level split is compiled to node \%split:[num\_users=3]=call\_function[target=torch.functional.split](args(\%l\_x\_, [4, 4, 24]), kwargs=\{dim: 1\}), it shows that the input of the split node is \%l\_x\_, and it has three users, representing item 0, item 1, and item 2 along the dim=1, each of them has the size 4, 4, and 24 in dim 1.

To better interpret the Torch-IR graph produced by PT2 for the example code in Listing~\ref{lst:code}, we present its graph representation in Fig.~\ref{fig:split-cat}.
In essence, the example splits the input tensor $x$ into two components.
The last component is further divided into two parts.
Lastly, all the above pieces are concatenated into a single node, which serves as the final return value.
The first-level split is compiled into the node \textit{\%split}, where the input to the split operation is node \textit{\%l\_x\_}.
This \textit{\%split} node has two users \textit{\%items0} and \textit{\%item1} with respective sizes 2 and 4 along dimension 1.

\subsection{Representative Case Studies} \label{sec:cases}
% \subsection{Training Queries Per Second Improvement}
\textbf{1) Training Queries Per Second Improvement:} Training queries per second (QPS), which is defined as \textit{number of samples in a global batch / iteration time}, is a key metric to measure the throughput of machine learning model training.
QPS quantifies how many training samples are processed by the model per second during the training loop. This metric is crucial for understanding and optimizing training efficiency, resource utilization, and cost.
We next give an example on our real-word recommendation models on how the pattern of \textit{pad\_aten\_mm\_pass} helps to boost QPS.

\begin{figure*}[!t]
    \centering
    \begin{minipage}[t]{0.48\textwidth}
        \centering
    \includegraphics[width=1\linewidth]{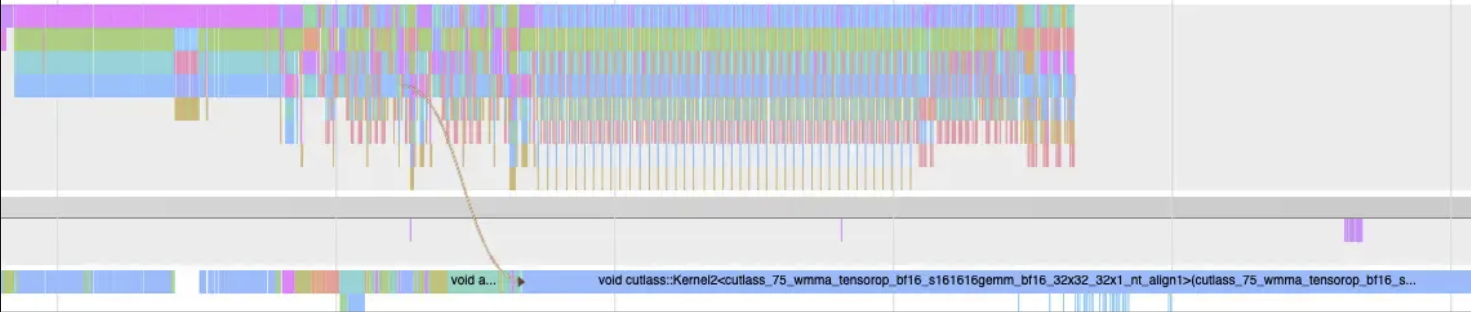}
    \caption{Long computation kernel \textit{cutlass\_75\_xxx}.}
    \label{fig:long_kernel}
    \end{minipage}
    \hfill
    \begin{minipage}[t]{0.48\textwidth}
        \centering
    \includegraphics[width=1\linewidth]{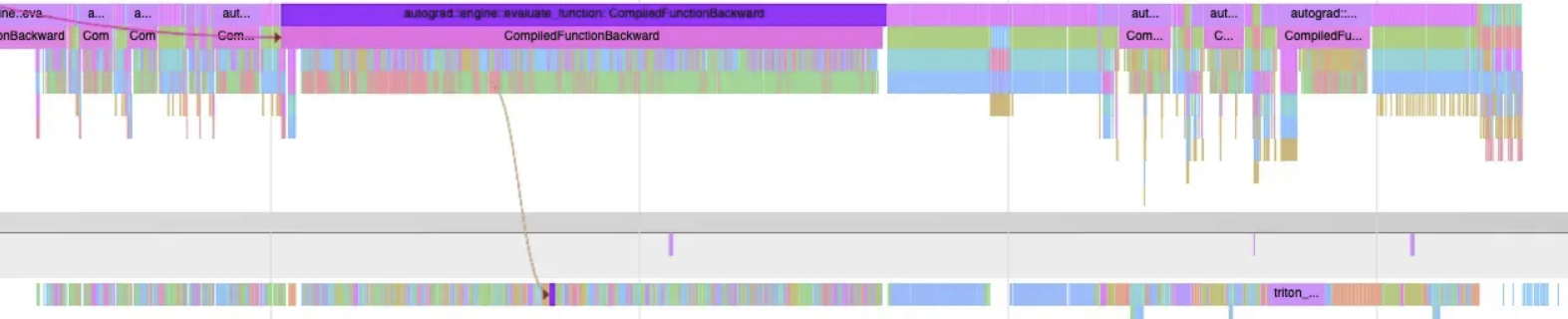}
    \caption{Long computation kernel disappears after padding.}
    \label{fig:after_pad}
    \end{minipage}
\end{figure*}

% \noindent\textbf{Example:}
In our experiments, we encountered a QPS performance bottleneck when enabling \textit{torch.bfloat16} data type for training using \textit{NVIDIA A100}.
Our GPU trace analysis revealed that the inefficiency stems from the \textit{aten.mm} operation in the backward pass, with a GEMM shape of $[M, K] \times [K, N] = [2, 15541760] \times [15541760, 13]$.
This GEMM shape is rather unusual due to its large $K$ value and misaligned $M$ and $N$ values (not multiples of 8).
As shown in Fig.~\ref{fig:long_kernel}, a kernel called \textit{cutlass\_75\_xxx} (blue long bar) was launched to process the \textit{aten.mm}, which proved to be highly inefficient. 

% \begin{figure}[t!]
%     \centering
%     \includegraphics[width=0.8\linewidth]{pictures/split_cat.png}
%     \caption{Graph representation of split-concat example.}
%     \label{fig:split-cat}
% \end{figure}

% \begin{figure}[t!]
%     \centering
%     \includegraphics[width=.85\linewidth]{pictures/pad_mm_pattern.png}
%     \caption{GEMM shape alignment padding.}
%     \label{fig:pad_mm_pattern}
% \end{figure}

To resolve the issue, we designed a new pattern to pad the matrices before they do the \textit{aten.mm} (see Fig.~\ref{fig:pad_mm_pattern}).
Specifically, we pad the shape of the matrix $m_2$ to be $[15541760, 16]$, ensuring that $N$ is a multiple of 8.

This alignment allows the operation to launch the more efficient kernel \textit{cutlass\_80\_xxx} (much smaller purple bar), eliminating the long-running kernel observed earlier (Fig.~\ref{fig:after_pad}).

To ensure semantic equivalence for outputs and gradients, we follow these three steps: i) {\em Padding:} before performing the matrix multiplications, the input tensors are padded so their dimensions $(M, K, N)$ are multiples of the kernel's preferred alignment (e.g., 8 for torch.bfloat16); ii) {\em Matmul:} The padded tensors are used for the computation, resulting in a larger output tensor; iii) {\em Slicing:} After matmul, the output is sliced to remove the padded regions, restoring the original shape and values.
% Thus, the padding/slicing process will {\em not} change the semantics.
\begin{figure*}[h]
    \centering
    % Left Subfigure - Prior to Decomposition Pass
    \begin{subfigure}[b]{0.48\textwidth}
        \centering
        \includegraphics[width=\linewidth]{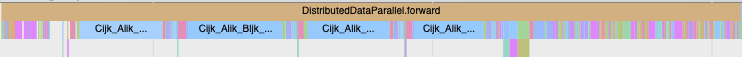}
        \caption{Prior to decomposition pass.}
        \label{fig:before_decompose}
    \end{subfigure}
    \hfill % Pushes the figures out to leave a clean gutter space
    % Right Subfigure - After Decomposition Pass
    \begin{subfigure}[b]{0.48\textwidth}
        \centering
        \includegraphics[width=\linewidth]{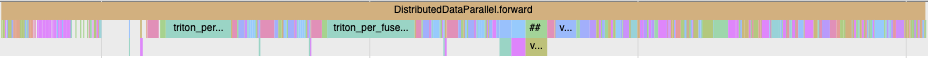}
        \caption{After decomposition pass.}
        \label{fig:after_decompose}
    \end{subfigure}
    
    % Main overarching caption
    \caption{GPU execution trace profile comparison: (a) Long-running GEMM kernels prior to applying the \textit{decompose\_mm\_pass}; (b) Execution timeline after applying the optimization pass, demonstrating the elimination of blocking GEMM kernels.}
    \label{fig:gemm_decomposition_traces}
\end{figure*}

\begin{figure*}[h]
    \centering
    \begin{minipage}[t]{0.48\textwidth}
        \centering
        \includegraphics[width=1\linewidth]{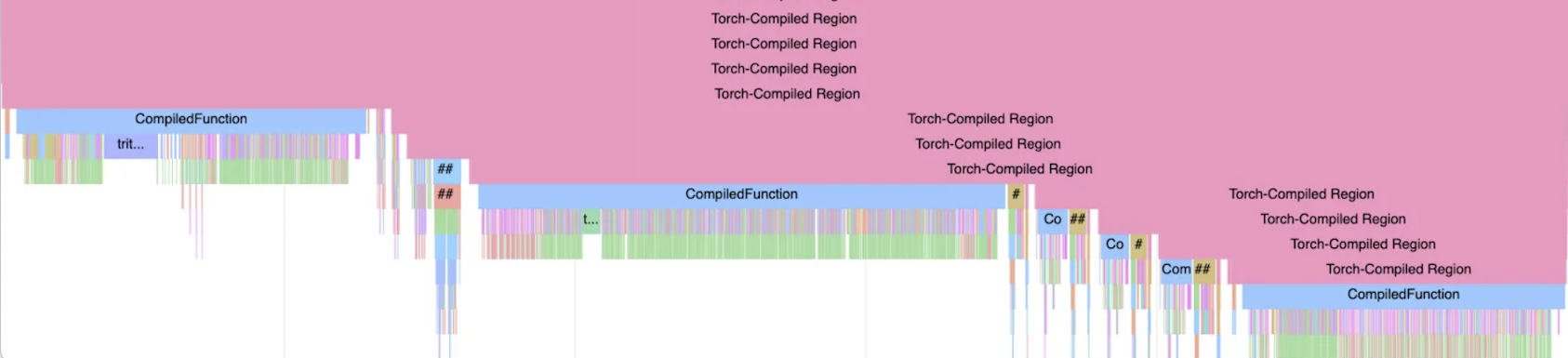}
        \caption{GPU trace screenshot without split cat patterns.}
        \label{fig:before_compile}
    \end{minipage}
    \hfill
    \begin{minipage}[t]{0.48\textwidth}
        \centering
        \includegraphics[width=1\linewidth]{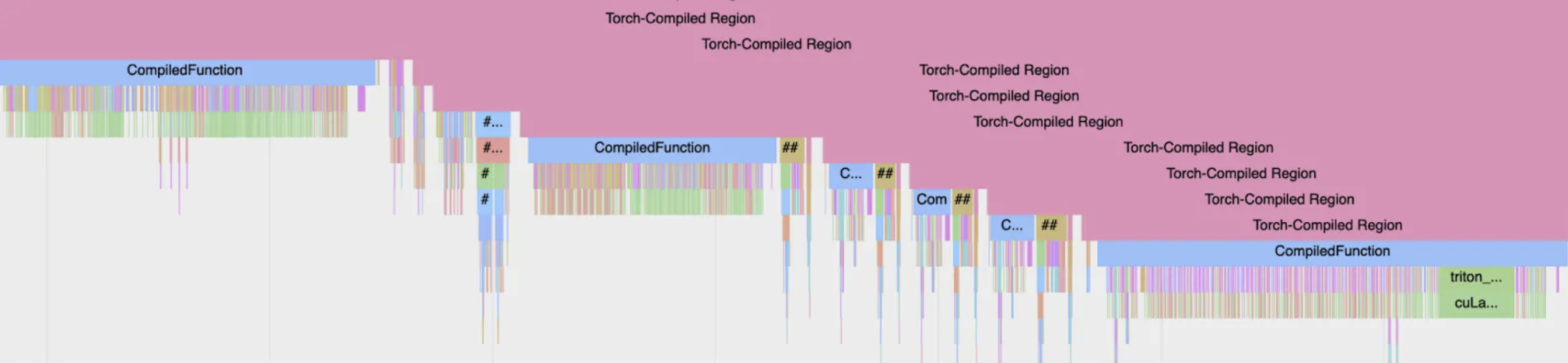}
        \caption{GPU trace screenshot with split cat patterns.}
        \label{fig:after_compile}
    \end{minipage}
\end{figure*}

\begin{figure}[h]
    \centering
    \includegraphics[width=.8\linewidth]{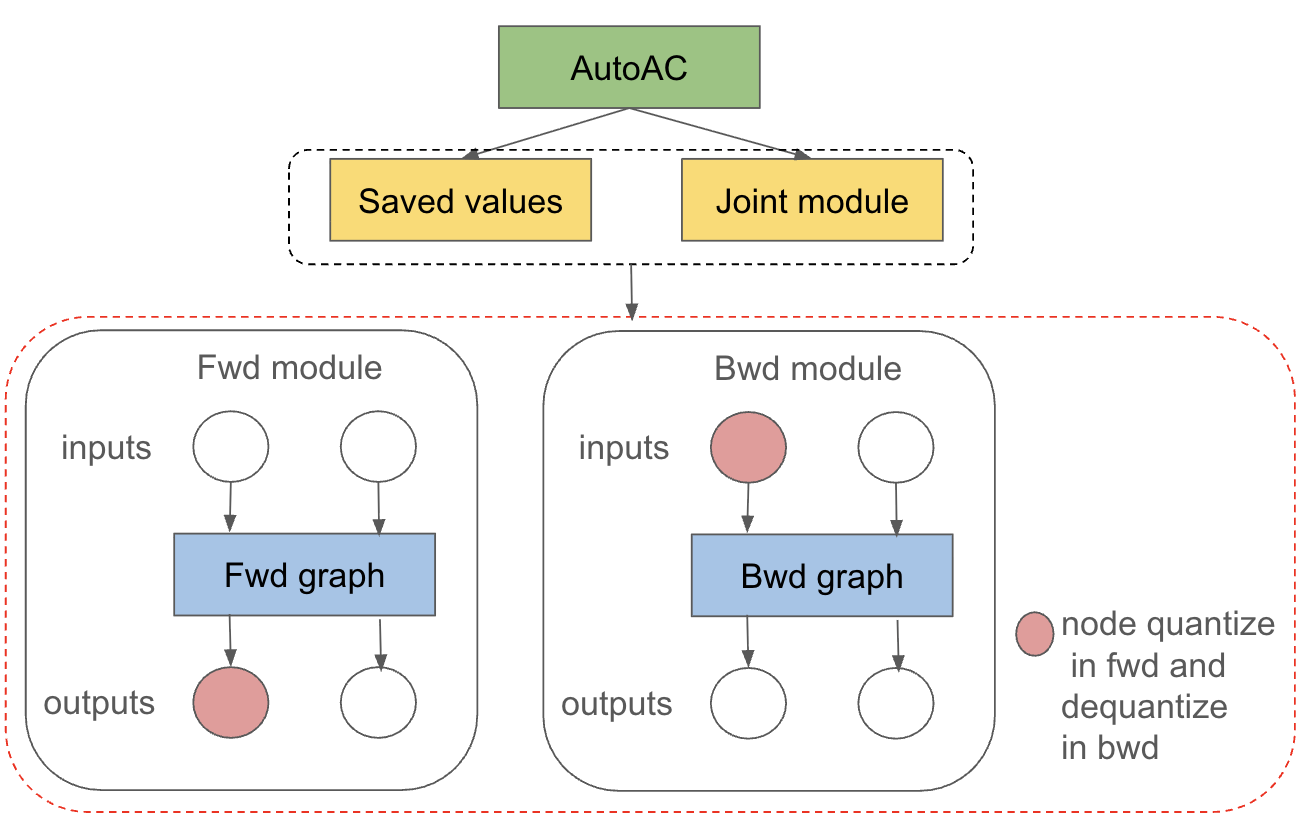}
    \caption{Overview of activation quantization pattern.}
    \label{fig:fp8_pattern}
\end{figure}

\textbf{2) decompose mm pass}:  We observe that computationally expensive, memory-bound matrix multiplication operators (\textit{bmm}, \textit{addmm}, and \textit{mm}) can be accelerated by decomposing the multiplication into pointwise operations followed by reductions along the K dimension.  
When combined with PT 2.x, which supports vertical fusion of pointwise operators, this decomposition enables execution within a single fused kernel, significantly improving efficiency.  
This approach is particularly effective when the matrix multiplication involves a very large M dimension and a small K dimension.  

% \begin{figure}[ht]
%     \centering
%     \includegraphics[width=1\linewidth]{pictures/before_decompose.png}
%     \caption{GPU trace of long-running GEMM kernels prior to applying the \textit{decompose\_mm\_pass}.}
%     \label{fig:before_decompose}
% \end{figure}

As an illustrative case, consider model type B, which executes four \textit{aten.mm} kernels with shapes  
[18583840, 8] $\times$ [8, 8],  
[25247936, 4] $\times$ [4, 2],  
[22502112, 8] $\times$ [8, 8], and  
[17980096, 4] $\times$ [4, 2].  
These kernels exhibit long runtimes, each exceeding 914 ms.  
As shown in Fig.~\ref{fig:before_decompose}, the GPU trace contains four distinct long-duration kernels (blue boxes labeled with “Cjik”), indicating significant bottlenecks.  

% \begin{figure}[ht]
%     \centering
%     \includegraphics[width=1\linewidth]{pictures/after_decompose.png}
%     \caption{GPU trace after applying the \textit{decompose\_mm\_pass}, eliminating long GEMM kernels.}
%     \label{fig:after_decompose}
% \end{figure}

After applying the \textit{decompose\_mm\_pass}, these long GEMM kernels are eliminated.  
The decomposed pointwise operations are subsequently fused by PT 2.x into more efficient kernels.  
As a result, the forward latency decreases from 923 ms to 476 ms, yielding a substantial QPS improvement. 

\textbf{3) Peak Memory Usage Reduction:} As the size of large models continues to grow, memory capacity becomes a critical bottleneck. 
Issues such as Out-of-Memory (OOM) error becomes more pronounced when batch sizes are increased or models are further scaled. 
For massive models such as LLaMA~\cite{touvron2023llama}, optimizing memory consumption is essential not only for training scalability but also for overall GPU resource efficiency.

Among the various contributors to memory overhead, activations account for a substantial proportion. 
Thus, optimizing activation storage and management is central to scaling large models effectively. 
Activation quantization~\cite{czako2025addressing} has emerged as a key technique to address this challenge by reducing memory requirements while maintaining model fidelity. 
By quantizing activation tensors, we can significantly lower memory consumption, thereby allowing larger models to be trained and deployed on existing hardware. 
This directly alleviates GPU memory pressure and enables more efficient and cost-effective scaling.

We use Fig.~\ref{fig:fp8_pattern} to illustrate the high-level concept of the \textit{activation\_quantization\_aten\_pass}. 
This pattern is built on top of AutoAC that determines which activations need to be preserved for backward computation. 
Based on the pattern configuration, we then select activations from the AutoAC~\cite{autoac2025} saved tensors to be quantized into a lower-precision format, such as FP8, in the forward pass to reduce the overall memory footprint. 
During the backward pass, the tensors are dequantized back to their original precision before use to minimize the precision impact.

% To demonstrate the effectiveness of activation quantization, we consider a simplified model consisting of five graphs in the forward pass. 
% In the baseline case, activations are stored in BF16, while in the optimized case, they are quantized to FP8. 
% From the memory snapshots, we observe that quantization can reduce the activation footprint by 5 GB.

% As shown in Fig.~\ref{fig:fp8_example}, quantized tensors that need to be used in the backward pass, created in the forward pass, occupy only half the memory of their BF16 counterparts.
% The BF16 tensors are properly released after their use in graph~b (highlighted in the green box). 
% However, in the red box, we observe that BF16 tensors used only in graph~d are not released until after graph~e, revealing an opportunity for further optimization in PyTorch’s memory management. 
% Addressing this inefficiency would enable additional reductions in peak memory usage.

\textbf{4) Compile Time Reduction:} Compiling very large models can take a long time (e.g., an hour or more) on cold starts, particularly for recommendation models with complex architectures beyond Transformers. 
Reducing compile time is therefore essential for lowering infrastructure costs~\cite{compiletime2025}. 
\alg addresses this challenge by eliminating excessive numbers of small Triton kernels, thereby improving compilation efficiency. 
As an example, we show that the use of split-concat patterns results in a compile-time reduction exceeding 400 seconds.

In the baseline case, the model exhibits a compile time exceeding 4000 seconds. 
As shown in Fig.~\ref{fig:before_compile}, the GPU trace reveals a large number of small triton kernels, which contribute significantly to compile time overhead (large green part).  
By applying three \alg split cat optimization patterns to the model, namely, \textit{split\_stack\_to\_cats\_pass}, \textit{split\_cat\_to\_slices\_pass}, and \textit{unbind\_cat\_to\_view\_pass}, the number of \textit{triton\_poi\_fused\_stack\_xxx} and \textit{triton\_poi\_fused\_cat\_xxx} kernels is substantially reduced, as shown in Fig.~\ref{fig:after_compile}. 
This reduction is also visible in the trace comparison, where the green regions (corresponding to split cat operations) shrink significantly once the new patterns are applied.  
These results confirm that \alg effectively eliminates redundant Triton kernels, thereby reducing compile time and improving efficiency.  

% \begin{figure}
%     \centering
%     \includegraphics[width=1\linewidth]{pictures/after_compile.png}
%     \caption{GPU trace screenshot with split--cat patterns.}
%     \label{fig:after_compile}
% \end{figure}

\bibliographystyle{ACM-Reference-Format}
\bibliography{sample-base}
\appendix

\end{document}